\documentclass[preprint,showpacs,preprintnumbers,amsmath,amssymb]{revtex4}

\usepackage{graphicx}% Include figure files
\usepackage{dcolumn}% Align table columns on decimal point
\usepackage{bm}% bold math
\usepackage{color}
\usepackage[latin1]{inputenc}

\newcommand{\ud}[1]{{#1^{\dagger}}}

\newcommand{\ket}[1]{\left| #1\right\rangle}

\newcommand{\mean}[1]{\langle#1\rangle}
\newcommand\Tr{\mathrm{Tr}}

\begin{document}

%\preprint{APS/123-QED}

\title{Luminescence Spectra of Quantum Dots in Microcavities. I.~Bosons.}

\author{Fabrice P.~Laussy}
\author{Elena del Valle}
\author{Carlos Tejedor}
\affiliation{%
Departamento de F\'{\i}sica Te\'orica de la Materia
Condensada, Universidad  Aut\'onoma de Madrid, Spain, fabrice.laussy@uam.es
}%

\date{\today}% It is always \today, today,

\begin{abstract}
  We provide a unified theory of luminescence spectra of coupled
  light-matter systems realized with semiconductor heterostructures in
  microcavities, encompassing: $i)$~the spontaneous emission case,
  where the system decays from a prepared (typically pure) initial
  state, and $ii)$~luminescence in the presence of a continuous,
  incoherent pump. While the former case has been amply discussed in
  the literature (albeit mainly for the case of resonance), no
  consideration has been given to the influence of the incoherent
  pump. We show how, by provoking a self-consistent quantum state, the
  pump considerably alters the emission spectra, even at vanishing
  intensities. The main outcome of our analysis is to unambiguously
  identify strong-coupling in situations where it appears in disguise
  or only seems to appear. Here, we consider bosonic matter fields, in
  which case fully analytical solutions can be obtained.  This
  describes the case of quantum wells or large quantum dots, or the
  limit of low excitation where the average populations remain much
  smaller than one.
\end{abstract}

\maketitle


\section{Introduction}

The dynamics of an optical emitter changes drastically when it is
placed in a cavity. The cavity alters the density of states of optical
modes, and therefore increases or inhibits interactions with the
emitter.  The effect was first put to use by Purcell in nuclear
magnetic resonance for the practical purpose of thermalizing spins at
radio frequencies, by bringing down their relaxation time
from~$\approx10^{21}$s to a few minutes~\cite{purcell46b}. Kleppner
applied the same idea in the opposite way, to increase the relaxation
time of an excited atom, i.e., to inhibit its spontaneous emission
(SE)~\cite{kleppner81a}.  The emitter, that in the case of Purcell was
sought to be resonant with the cavity mode to increase the photon
density of states with respect to the vacuum, was in the case of
Kleppner put out of resonance, namely in a photonic gap, where the
photon density of states is smaller than in vacuum. This tuning of the
relaxation time of an emitter placed in a cavity, now known as the
\emph{Purcell effect}, has many potential technological applications,
one of the most compelling being the decrease of the lasing
threshold. The effect, that had first been actively looked for with
atoms in cavities~\cite{goy83a}, was therefore also intensively (and
more recently) pursued in the solid state, more prone for massive
technological implementations.  Semiconductor heterostructures are the
state of the art arena for this purpose. They allow to engineer, with
an ever rising control, the solid state counterpart of the atomic
system to match or isolate their excitation spectra and thus control
their behaviour.  Typical examples are quantum dots (QDs) placed in
cavities made in micropillars, microdisks or photonic crystals, where
Purcell inhibition has been distinctly
demonstrated~\cite{gerard98a,kiraz02a}.  In this paper, we shall be
concerned chiefly with the observed spectra of emission of such an
emitter placed in a cavity. In the regime where the effect of the
cavity is to lengthen or shorten the lifetime of the excitation, the
consequence in the optical spectra is to narrow or broaden the line,
respectively.

In the description of the Purcell effect, the possible reabsorption of
the photon by the emitter is so weak that it can be neglected.  It is
responsible for the energy shift known as the Lamb shift, that, in
quantum electrodynamics, is interpreted as the perturbative influence
of virtual photons emitted and re-absorbed by the emitter. In the case
of inhibition of the spontaneous emission, this shift is indeed orders
of magnitude smaller than the radiative broadening.  In the case where
emission is enhanced, and the linewidth narrowed, the probability of
reabsorption of a photon by the emitter becomes closer to that of
escaping the cavity, until the perturbative---so-called
\emph{weak-coupling} (WC)---regime breaks down and instead
\emph{strong coupling} (SC) takes place. In this case, photons emitted
are then reflected by the mirrors and there is a higher probability
for their reabsorption by the atom than for their leaking out of the
cavity.  A whole sequence of absorptions and emissions can therefore
take place, known as \emph{Rabi oscillations}.  This regime is of
greater interest, as it gives rise to new quantum states of the
light-matter coupled system, usually referred to as \emph{dressed
  states} in atomic physics and as \emph{polaritons} in solid-state
physics. Experimentally, SC is more difficult to reach, as it
requires a fine control of the quantum coupling between the bare
modes and in particular to reduce as much as possible all the
sources of dissipation. Theoretically, it is better dealt with by
first getting rid of the dissipation, and starting with the
\emph{strong-coupling Hamiltonian} ($\hbar $ is taken as 1 along the
paper):
\begin{equation}
  \label{eq:FriMar14013046UTC2008}
  H=\omega_a\ud{a}a+\omega_b\ud{b}b+g(\ud{a}b+a\ud{b})
\end{equation}
where $a$ and~$b$ are the cavity photon and material excitation field
operators, respectively, with bare mode energies~$\omega_a$
and~$\omega_b$, coupled linearly with strength~$g$.  The photon
operator is a Bose annihilation operator, satisfying the usual
commutation rule~$[a,\ud{a}]=1$.  Depending on the model for the
material excitation, $b$ is described by, typically, another harmonic
oscillator (linear model~\cite{hopfield58a}) or a two-level system
(Jaynes-Cummings model~\cite{jaynes63a}). Those are the most
fundamental cases as they describe material fields with Bose and Fermi
statistics, respectively.  Possible extensions are a collection of
harmonic oscillators~\cite{rudin99a} or of two-level
systems~\cite{dicke54a}, a three-level system~\cite{bienert04a}, etc.
This paper contains the first part of our work in which we address
exclusively the case where~$b$ also follows Bose statistics.  This is
an important case for two reasons.  The first one is that in many
relevant cases, the matter-field is indeed bosonic, such as the case
of quantum wells, or large quantum dots, at low density of
excitations. The second reason is that this case provides the limit
for vanishing excitations of all the other cases, and is fully
solvable analytically. As such, this case provides the backbone for
the qualitative understanding of SC. We leave for a further work the
case of fermions at larger pumping, more relevant when dealing with
small QDs that confine excitations, and more prone to involve genuine
quantum mechanics as one quantum of excitation can alter the system's
response. The drawback is that numerical computation is required in
this case, and the discussion is therefore of a somehow less
fundamental character.

For the rest of the text, it is therefore understood that $b$ is also
a Bose operator. For convenience, we shall refer to it as the
\emph{exciton} operator, after the name of a bound electron-hole pair
in a semiconductor. Likewise, we shall prefer such terminology as a
Quantum Dot (QD) rather than an atom, or polaritons, rather than
dressed states, etc. For most purpose, this is semantics only and the
results apply in a wide range of systems.

Equation~(\ref{eq:FriMar14013046UTC2008}) can be straightforwardly
diagonalized for bosonic modes,
giving~$H=\omega_\mathrm{U}\ud{p}p+\omega_\mathrm{L}\ud{q}q$, where
\begin{equation}
  \label{eq:SunMar30211523UTC2008}
  \omega_{\substack{\mathrm{U}\\\mathrm{L}}}=\frac{\omega_a+\omega_b}2\pm\sqrt{g^2+\frac{\Delta^2}4}\,,
\end{equation}
with new Bose operators $p=(\cos\theta ) a+(\sin\theta ) b$
and~$q=-(\sin\theta) a+(\cos\theta) b$, determined by the \emph{mixing
  angle},
$\theta=\arctan\big((\frac\Delta2+\sqrt{g^2+\frac{\Delta^2}4})/g\big)$,
and the detuning:
% with~$\theta=(\Delta+\sqrt{4g^2+\Delta^2})/(8g^2+2\Delta^2+2\Delta\sqrt{4g^2+\Delta^2})$
% the so-called \emph{mixing angle}, chiefly determined by the
% detuning~$\Delta$ between the bare modes:
%
\begin{equation}
  \label{eq:SunMar30210322UTC2008}
  \Delta=\omega_a-\omega_b\,.
\end{equation}
These new modes are the polaritons (or dressed states) with quantum
states~$\ket{\mathrm{U}}=\ud{p}\ket{\mathrm{vac}}=\cos\theta\ket{1,0}+\sin\theta\ket{0,1}$
and~$\ket{\mathrm{L}}=\ud{q}\ket{\mathrm{vac}}=-\sin\theta\ket{1,0}+\cos\theta\ket{0,1}$,
where~$\ket{\mathrm{vac}}$ is the vacuum, $\ket{1,0}$ is the Fock
state of one photon and~$\ket{0,1}$ the Fock state of one exciton.

\begin{figure}[tp]
  \centering
  \includegraphics[width=.66\linewidth]{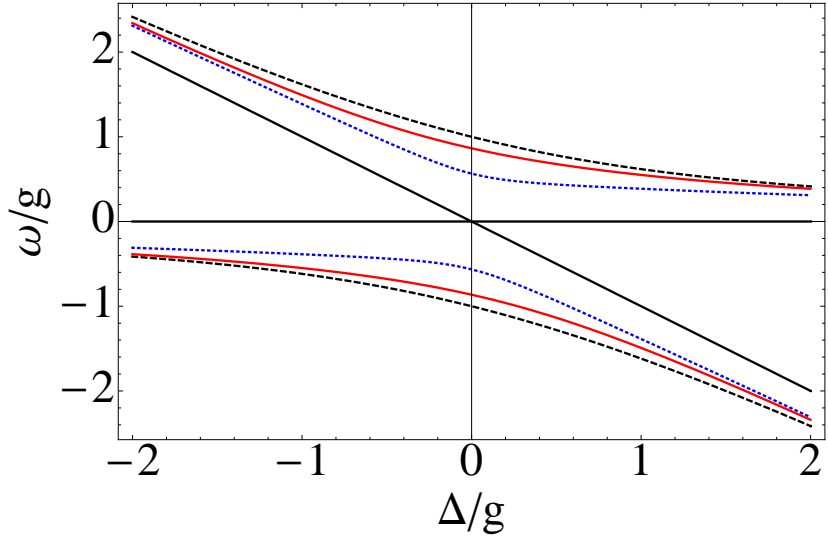}
  \caption{(Color online) Solid black: Bare energies of the cavity
    photon (horizontal line) and of the exciton (tilted) as a function
    of detuning $\Delta$. Dashed black: Eigenenergies of the system
    Hamiltonian, without dissipation nor pumping
    [eqn~(\ref{eq:SunMar30211523UTC2008})].  The exciton-like state at
    large negative $\Delta$ has become a photon-like state at large
    positive $\Delta$, and vice-versa.  Around $\Delta=0$, both modes
    are an admixture of exciton and photon. Dotted blue: Correction of
    the eigenenergies when pump and decay are taken into account for
    parameters of point~(c) in Fig.~\ref{fig:FriMay16003338UTC2008}.
    Solid red: Actual position of the observed peaks in the PL spectra
    for the same parameters. For these parameters, the three
    descriptions of SC give the same qualitative results.}
  \label{fig:SunMar30215102UTC2008}
\end{figure}

The energies defined by eqn~(\ref{eq:SunMar30211523UTC2008}) are
displayed in Fig.~\ref{fig:SunMar30215102UTC2008} with dashed lines,
on top of that of the bare modes, with thick lines, as detuning is
varied by changing the energy of the emitter and keeping that of the
cavity constant.  The \emph{anticrossing} always keeps the upper
mode~U higher in energy than the lower~L one, strongly admixing the
light and matter character of both particles. At resonance, this
mixing is maximal:
\begin{equation}
  \label{eq:ThuApr3114803CEST2008}
  \ket{\mathrm{\substack{\mathrm{U}\\\mathrm{L}}}}=\frac1{\sqrt2}\left(\ket{1,0}\pm\ket{0,1}\right)\,.
\end{equation}
If the system is initially prepared as a bare state---which is the
natural picture when reaching the SC from the excited state of an
emitter---the dynamics is that of an oscillatory transfer of energy
between light and matter. In an empty cavity, the time evolution of a
state which was an exciton at~$t=0$, is given by:
\begin{subequations}
  \begin{align}
    \label{eq:ThuApr3121629CEST2008}
    \ket{\psi(t)}&=\exp\big(-iHt\big)\ket{0,1}\,,\\
    &=\sin\theta\cos\theta(e^{-i\omega_\mathrm{U} t}-e^{-i\omega_\mathrm{L}t}) \ket{1,0}\nonumber \\
    &+(\sin^2\theta e^{-i\omega_\mathrm{U} t}+\cos^2 \theta
    e^{-i\omega_\mathrm{L}t})\ket{0,1}\,,
  \end{align}
\end{subequations}
which results in oscillations between the bare modes at the so-called
\emph{Rabi frequency}:
\begin{equation}
  \label{eq:ThuJun5115514BST2008}
  \mathcal{R}=\frac{\omega_\mathrm{U}-\omega_\mathrm{L}}2=\sqrt{g^2+\frac{\Delta^2}4}\,.
\end{equation}

The \emph{emission spectrum} of the system requires dissipation, as it
is an obvious practical requisite that the excitation should
eventually leak out of the system to be detected from the outside [the
\emph{energy spectrum} of the system is given by
eqn~(\ref{eq:SunMar30211523UTC2008})].  Dissipation is intrinsic to
the bare modes: both the cavity photon and the exciton have a finite
lifetime. In presence of dissipation, the system is upgraded from a
Hamiltonian description, eqn~(\ref{eq:FriMar14013046UTC2008}), to a
Liouvillian description, with a quantum dissipative master equation
for the density matrix~$\rho$ defined in the tensor product of the
light and matter Hilbert spaces~$\mathcal{H}_a$
and~$\mathcal{H}_b$~\cite{carmichael_book02a}:
\begin{equation}
  \label{eq:FriMar14013423UTC2008}
  \partial_t\rho=\mathcal{L}_\gamma\rho\,.
\end{equation}
The Liouvillian~$\mathcal{L}_\gamma$ still contains the Hamiltonian
dynamics of SC, but also takes into account the decays of both the
cavity and the emitter, with rates~$\gamma_a$ and~$\gamma_b$,
respectively:
\begin{equation}
  \label{eq:ThuOct18162449UTC2007}
  \mathcal{L}_\gamma \rho=i[\rho,H]+\sum_{c=a,b}\frac{\gamma_c}2(2c\rho\ud{c}-\ud{c}c\rho-\rho\ud{c}c)\,,
\end{equation}
in which we have considered temperature equal to zero. This equation
has been extensively studied~\cite{carmichael89a}, although, to the
best of our knowledge, not in its most general form.  The typical
restrictions have been to consider the case of resonance,
$\omega_a=\omega_b$, with only one particular initial condition,
namely, the excited state of the emitter in an empty cavity, and to
detect the emission of the emitter itself. All together, they describe
the spontaneous emission of an emitter placed into a cavity with which
it enters into SC. This has been the topical case for decades as this
was the case of experimental interest with atoms in cavities.

With the advent of SC in other systems, other configurations start to
be of interest. With a QD in a microcavity, the
\emph{detuning}~$\Delta$ between the modes,
eqn~(\ref{eq:SunMar30210322UTC2008}), is a crucial experimental
parameter, as it can be easily tuned and to a great extent, for
instance by applying a magnetic field or changing the temperature.
Also in this case, the detection is in the optical mode of the cavity,
rather than the direct emission of the exciton emission, because the
latter is awkward for various technical reasons of a more or less
fundamental character (an example of a fundamental complication is
that the emission is enhanced in the cavity mode and suppressed
otherwise, and the exciton lifetime is typically much longer, so the
exciton emission is much weaker; an example of a petty technical
complication is that the exciton detection should be made at an angle
and, practically, a lot of samples are grown on the same substrate.
Both the substrates and other samples hinder the lateral access to one
given sample, whereas all are equally accessible from above). In our
system where both modes are bosonic, symmetry allows us to focus on
the cavity emission without loss of generality, as we can obtain the
leaky excitonic emission by simply exchanging indexes $a,b$ (the
spectrum could also have photon-exciton crossed terms that could be
computed in a similar way).  When we shall turn to the case of a
fermionic matter-field, where the exciton emission will become
distinctly different and for that reason, important, we shall address
exciton emission separately~\cite{psip_delvalle08a}.

Regarding the initial condition, more general quantum states can now
be realized, at least in principle, by coherent control, pulse shaping
or similar techniques. Additionally and more importantly, the type of
excitation of a cavity-emitter system in a semiconductor is of a
different character: the excitation is typically injected by either a
continuous wave (cw) laser far above resonance, creating electron-hole
pairs that relax incoherently to excite the QD in a continuous flow of
excitations, or by electrical pumping as in lasers. This pumping, that
is of an incoherent nature typical of semiconductor physics, brings
many fundamental changes into the problem that go beyond the mere
generalisation of eqn~(\ref{eq:ThuOct18162449UTC2007}). Among the most
obvious ones, let us already mention that pure states of the like of
eqn~(\ref{eq:ThuApr3114803CEST2008}) do not correspond to the
experimental reality. Instead, the system is maintained in a partially
mixed state with probabilities~$p(n)$ to realize the~$n$th excited
state. In all cases, a steady state is imposed by the interplay of
pumping and decay. The Rabi oscillations of the populations---that is,
the coherent exchange of energy between the modes---are always washed
out, regardless of the photon-like, exciton-like or polariton-like
(eigenstate) character of the density matrix.

In this work, we address both the emission spectra obtained in a
configuration of spontaneous emission (SE)---where an initial state is
prepared and left to decay, as ruled by
eqns~(\ref{eq:FriMar14013423UTC2008}--\ref{eq:ThuOct18162449UTC2007})---under
its most general setting, and the case of luminescence emission under
the action of a continuous and incoherent pumping that establishes a
steady state (SS). We bring all results under a common and unified
formalism and show how none of the cases fully encompasses the
other. The rest of the paper is organised as follows.  In
Section~\ref{sec:ThuApr3131307CEST2008}, we present the complete model
and we derive and discuss its equation of motion, focusing on the
single-time dynamics.  In Section~\ref{sec:ThuApr3131407CEST2008}, we
obtain fully analytically the main results in both of the cases
explicated above, this time focusing more on the two-time dynamics,
which Fourier transform gives the luminescence spectra. In
Section~\ref{sec:ThuApr3161508CEST2008}, we discuss the mathematical
results derived in the two previous sections, accentuating the
physical picture and relying on particular cases for illustration.  We
consider, in this Section, specifically the case of resonance, where
all the concepts manifest more clearly. Finally, in
Section~\ref{sec:ThuApr3161611CEST2008}, we give a summary of the main
results and provide an index of all the important formulas and key
figures of this text. We conclude with a short overview of the
continuation of this work that replaces the bosonic emitter with a
fermionic one.

\section{Model}
\label{sec:ThuApr3131307CEST2008}

\begin{figure}[tbhp]
  \centering
  \includegraphics[width=.6\linewidth]{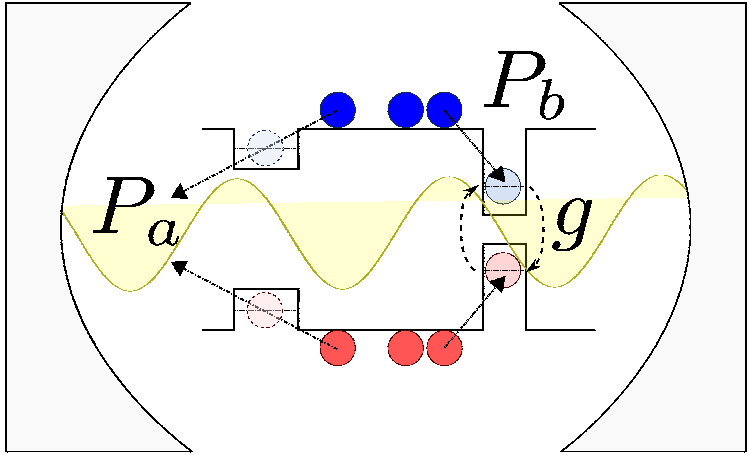}
  \caption{(Color online) Schema of our system for the SS case:
    self-assembled QDs in a semiconductor microcavity. The QD sketched
    on the right is in SC with the cavity mode with coupling
    strength~$g$, while the one of the left is in WC. The
    electron-hole pairs created by the incoherent pumping of the
    structure provide an effective electronic pumping, $P_b$, of the
    SC dot, while the pumping of the WC dot results in an effective
    cavity pumping~$P_a$, through rapid conversion of the excitons
    into cavity photons.}
  \label{fig:FriMar14002142UTC2008}
\end{figure}

Our model for the coupling of a QD with a single cavity mode in the
presence of incoherent and continuous pumping is sketched in
Fig.~\ref{fig:FriMar14002142UTC2008}. One QD is strongly coupled to
the cavity mode with interaction strength~$g$, and is continuously,
and incoherently, excited by an electronic pumping~$P_b$, which, in
the microscopic picture, is linked to the rate at which electron-hole
pairs relax into the dot. This rate is related in some way to the
pumping exerted by the experimentalist. Because of the incoherent
nature of this pumping, together with that of the damping, the
off-diagonal elements of the QD reduced density matrix (that hold the
coherence) are washed out in the SS. We also consider another type of
pumping, $P_a$, that offers a counterpart for the cavity by injecting
photons incoherently at this rate. The major factor to account for
such a term is the presence of many other QDs, that have been grown
along with the one of interest. Those only interact weakly with the
cavity. In most experimental situations so far, it is indeed difficult
to find one dot with a sufficient coupling to enter the
non-perturbative regime. When this is the case, all the other dots
that remain in WC become \emph{spectators} of the SC physics between
the interesting dot and the cavity, and their presence is noticed by
weak emission lines in the luminescence spectrum and an increased
cavity emission. They are also excited by the electronic pumping that
is imposed by the experimentalist, but instead of undergoing SC, they
relax their energy into the cavity by Purcell enhancement or
inhibition, depending on their proximity with the cavity mode. This,
in turn, results in an effective pumping of the
cavity~\cite{psip_laussy08a}.

To model these two continuous and incoherent pumping, we use Lindblad
terms $\mathcal{L}_\mathrm{P}$ that substitute the annihilation
operator $c$ (standing for~$a$ or~$b$) by the creation one
$c^\dagger$, and vice-versa, a procedure which is known to describe
pumping terms in a quantum rate equation~\cite{rubo03a}. The full
Liouvillian now reads (again at zero temperature):
\begin{subequations}
  \label{eq:L2}
  \begin{align}
    \partial _t \rho=&\mathcal{L}\rho=(\mathcal{L}_\gamma+\mathcal{L}_\mathrm{P})\rho\\
    =&i\omega_a[\rho,\ud{a}a]+i\omega_b[\rho,\ud{b}b]+ig[\rho,\ud{a}b+a\ud{b}]\\
    +&\sum_{c=a,b}\frac{\gamma_c}2(2c\rho\ud{c}-\ud{c}c\rho-\rho\ud{c}c)\\
    +&\sum_{c=a,b}\frac{P_c}2(2\ud{c}\rho c-c\ud{c}\rho-\rho c\ud{c})\,.\label{eq:ThuJun5131626BST2008}
  \end{align}
\end{subequations}

The microscopic derivation of line~(\ref{eq:ThuJun5131626BST2008})
follows from the usual Born-Markov
approximation~\cite{carmichael_book02a}. The case of electronic
pumping, for instance, is similar to the process of laser gain: the
medium requires an inversion of electron-hole population, something
that cannot be achieved by means of a simple harmonic oscillator heat
bath. The actual process of gaining an exciton in the QD involves the
annihilation an electron-hole pair in an external reservoir out of
equilibrium (representing either electrical injection or the capture
of excitons optically created at frequencies larger than the ones of
our system) and the emission of a phonon, that carries the excess of
energy, to another one (which can be in thermal equilibrium). A simple
effective description of this non-equilibrium process can be made by
an inverted harmonic oscillator with levels $E_p=-
\omega_\mathrm{P}(p+1/2)$ maintained at a \emph{negative
  temperature}~\cite{psip_gardiner91a}.  Since the raising operator
for the energy decreases the number of quanta of this oscillator, the
role of creation and destruction operators is indeed reversed with
respect to the usual case of damping.

For the sake of generality, we introduce \emph{effective broadenings},
that reduce to the decay rates in the SE case but get renormalized by
the pumping rate in the SS case:
\begin{subequations}
  \label{eq:ThuApr3180040CEST2008}
  \begin{align}
    \Gamma_{a,b}&=\gamma_{a,b}\,,&\text{(SE case)}\\
    \Gamma_{a,b}&=\gamma_{a,b}-P_{a,b}\,.&\text{(SS case)}
  \end{align}
\end{subequations}
We shall also use thoroughly the combinations:
\begin{equation}
  \label{eq:M2}
  \gamma_{\pm}=\frac{\gamma_a\pm\gamma_b}{4}\quad\text{and}\quad\Gamma_{\pm}=\frac{\Gamma_a\pm\Gamma_b}{4}\,.
\end{equation}

The renormalization of the linewidth in the presence of the pumping
term is a bosonic effect. In the case of a single harmonic oscillator
driven by pump and decay, the emission spectrum is a Lorentzian
lineshape with Full-Width at Half Maximum (FWHM) given by
$\gamma-P=P/\langle n\rangle$. The linewidth narrows with the number
of particle, in a way reminiscent of the Schallow-Townes
effect~\cite{scully_book02a}.

From the relations $\langle O\rangle=\Tr(O\rho)$ and
$\partial_t\langle
O\rangle=\Tr(O\partial_t\rho)=\Tr(O\mathcal{L}\rho)$, we can obtain
from eqn~(\ref{eq:L2}) the single-time mean values of interest for
this problem, by solving the equation of motion of the coupled
system:
\begin{equation}
  \label{eq:M6}
  \partial_{t}
  \begin{pmatrix}
    n_{a}\\
    n_{b}\\
    n_{ba}\\
    n_{ab}
  \end{pmatrix}
  =
  \begin{pmatrix}
    P_{a}\\
    P_{b}\\
    0\\
    0
  \end{pmatrix}
  +
  \begin{pmatrix}
    -\Gamma_a & 0 & ig & -ig\\
    0 & -\Gamma_b & -ig & ig \\
    ig & -ig & -i\Delta-2\Gamma_+&0\\
    -ig & ig & 0 & i\Delta-2\Gamma_+
  \end{pmatrix}
  \begin{pmatrix}
    n_{a}\\
    n_{b}\\
    n_{ba}\\
    n_{ab}
  \end{pmatrix}\,,
\end{equation}
where $n_{c}=\mean{\ud{c}c}\in\mathbb{R}$ (for $c=a,b$) and
$n_{ab}=\mean{\ud{a}b}=n_{ba}^*\in\mathbb{C}$. The SE case corresponds
to setting~$P_{a,b}=0$ and providing the initial conditions:
\begin{equation}
  \label{eq:FriApr4021006UTC2008}
  n_a^0\equiv n_{a}(0),\quad
  n_b^0\equiv n_{b}(0),\quad\text{and}\quad
  n_{ab}^0\equiv n_{ab}(0)\,.
\end{equation}
The solutions are heavy, but not devoid of interest.  For
completeness, and since this is a natural and potentially useful
result within the scope of this paper, we provide them in
Appendix~\ref{MonMay19214851CEST2008}.

On the other hand, the SS case corresponds to setting the time
derivative on the left hand side of eqn~(\ref{eq:M6}) to zero, and
solving the resulting set of linear equations. This yields:
\begin{subequations}
  \label{eq:WedMay28004656CEST2008}
  \begin{align}
    n_a^\mathrm{SS}&=\frac{g^2\Gamma_+(P_a+P_b)+P_a\Gamma_b(\Gamma_+^2+(\frac{\Delta}{2})^2)}{4g^2\Gamma_+^2+\Gamma_a\Gamma_b(\Gamma_+^2+(\frac{\Delta}{2})^2)}\,,\label{eq:MonMay19124027CEST2008}\\
    n_b^\mathrm{SS}&=\frac{g^2\Gamma_+(P_a+P_b)+P_b\Gamma_a(\Gamma_+^2+(\frac{\Delta}{2})^2)}{4g^2\Gamma_+^2+\Gamma_a\Gamma_b(\Gamma_+^2+(\frac{\Delta}{2})^2)}\,,\label{eq:SunMay25222529CEST2008}\\
    n_{ab}^\mathrm{SS}&=\frac{\frac{g}2(\gamma_aP_b-\gamma_bP_a)(i\Gamma_+-\frac{\Delta}2)}{4g^2\Gamma_+^2+\Gamma_a\Gamma_b(\Gamma_+^2+(\frac{\Delta}{2})^2)}\,.
  \end{align}
\end{subequations}

Both photonic and excitonic reduced density matrices are
diagonal. They correspond to thermal distributions of particles with
the above mean numbers~\cite{alicki88a}:
\begin{subequations}
  \begin{align}
    \label{eq:Solho}
    &\rho^\mathrm{a}_{n,p}=\sum_{m}\rho_{n,m;\,p,m}=\delta_{n,p}\frac{(n_a^\mathrm{SS})^n}{(1+n_a^\mathrm{SS})^{n+1}}\,,\\
    &\rho^\mathrm{b}_{m,q}=\sum_{n}\rho_{n,m;\,n,q}=\delta_{m,q}\frac{(n_b^\mathrm{SS})^m}{(1+n_b^\mathrm{SS})^{m+1}}\,.
  \end{align}
\end{subequations}

Behind their forbidding appearance,
eqns~(\ref{eq:WedMay28004656CEST2008}) enjoy a transparent physical
meaning, that they inherit from the semi-classical---and therefore
intuitive---picture of rate equations. For instance, when the coupling
strength between the two modes, $g$, vanishes, the solutions
$n_{a,b}=P_{a,b}/\Gamma_{a,b}$ are those of a source and sink problem
for bosons, i.e., of the kind $\partial_t n=-\gamma n+P(n+1)$,
featuring the famous Bose stimulation effect, whereby the probability
of relaxation towards the final state is increased by its population.

\section{Correlation functions and spectra}
\label{sec:ThuApr3131407CEST2008}

We now turn to the main goal of this paper, namely, the luminescence
spectrum of the system~$s(\omega)$. Physically, it is the intensity
(or mean number) of photons in the system with frequency~$\omega$,
i.e., by definition:
\begin{equation}
  \label{eq:ThuApr3171338CEST2008}
  s(\omega)=\langle\ud{a}(\omega)a(\omega)\rangle\,.\\
\end{equation}
This is proportional to the intensity of photons emitted by the cavity
at this frequency (the direct exciton emission from its
recombination is described in a similar way
by~$\langle\ud{b}(\omega)b(\omega)\rangle$). It will be more
convenient, throughout, to deal with normalized spectra:
\begin{equation}
  \label{eq:S3}
  S(\omega)=s(\omega)\Big/\int_0^\infty\langle\ud{a}a\rangle(t)dt\,,
\end{equation}
so that eqn~(\ref{eq:S3}) is now the density of probability that a
photon emitted by the system has frequency~$\omega$.  The Fourier
transform of~$a(\omega)$ relates the emission spectrum to a two-time
correlator through
$S(\omega)=\frac{1}{2\pi}\iint_0^{\infty}\langle\ud{a}(t_1)a(t_2)\rangle
e^{i\omega(t_2-t_1)}dt_1dt_2/\int\langle\ud{a}a\rangle(t)dt$, that
after a change of variables, can be expressed in terms of the
so-called first-order time autocorrelator:
\begin{equation}
  \label{eq:ThuApr3171846CEST2008}
  G^{(1)}(t,\tau)=\langle\ud{a}(t)a(t+\tau)\rangle
\end{equation}
for positive time delay $\tau=t_2-t_1>0$. All put together, this
yields the usual Fourier-pair relationship between the power spectrum
and the autocorrelation function:
\begin{equation}
  \label{eq:ThuApr3172523CEST2008}
  S(\omega)=\frac{1}{\pi\int_{0}^\infty\langle\ud{a}a\rangle(t)dt}\Re\int_0^\infty\int_0^\infty G^{(1)}(t,\tau)e^{i\omega\tau}\,d\tau dt\,.
\end{equation}
Equation~(\ref{eq:ThuApr3172523CEST2008}) holds as such in the SE
case. In the SS case, care must be taken with cancellation of
infinities brought by the ever-increasing time~$t$. A technical but
straightforward procedure (cf. Appendix~\ref{MonMay26103617CEST2008})
leads to the expression that explicitly gets rid of the
divergences---famously known as the Wiener-Khintchine
theorem~\cite{mandel95a}---that reads:
\begin{equation}
  \label{eq:MonMay19214206CEST2008}
  S^\mathrm{SS}(\omega)=\frac{1}{\pi}\frac{1}{n_a^\mathrm{SS}}\lim_{t\rightarrow\infty}\Re\int_{0}^{\infty}G^{(1)}(t,\tau)e^{i\omega\tau}d\tau\,.
\end{equation}
From now on, we shall refer with~``SE'' and~``SS'' to the expressions
that apply specifically to the spontaneous emission and to the steady
state, respectively, leaving free of index those that are of general
validity. In some cases, as for instance in
eqn~(\ref{eq:ThuApr3180040CEST2008}), no index is required if it is
understood that~$P_{a/b}$ are defined and equal to zero in the SE
case. For that reason, we shall leave~$\Gamma$ free of the SE/SS
redundant index.

To obtain the spectra of a system whose dynamics is dictated by
eqn~(\ref{eq:L2}), we therefore need to compute two-time dynamics.
This can be done thanks to the quantum regression theorem, according
to which, a set of operators~$A_{\{\alpha\}}$ that
satisfy~$\Tr(A_{\{\alpha\}}\mathcal{L}O)=\sum_{\{\beta\}}
M_{\{\alpha\beta\}}\Tr(A_{\{\beta\}}O)$ for
all~$O\in\mathcal{H}_a\otimes\mathcal{H}_b$ for
some~$M_{\{\alpha\beta\}}$, yields the equations of motion for the
two-time correlators as:
\begin{equation}
  \label{eq:ThuApr3174049CEST2008}
  \partial_\tau\langle
  O(t)A_{\{\alpha\}}(t+\tau)\rangle=\sum_{\{\beta\}}M_{\{\alpha\beta\}}\langle
  O(t)A_{\{\alpha\}}(t+\tau)\rangle\,.
\end{equation}
In the fully bosonic case, operators $a^n b^m$
with~$\{\alpha\}=(n,m)\in\mathbb{N}$ constitute such a set with~$M$
defined by
\begin{align}
    \label{eq:ruleslc}
    &M_{\substack{nm\\nm}}=-i(n\omega_a+m\omega_b)-n\frac{\Gamma_a}{2}-m\frac{\Gamma_b}{2}\, ,\\
    &M_{\substack{nm\\n+1,m-1}}=M_{\substack{mn\\m-1,n+1}}=-igm\, ,
\end{align}
and zero everywhere else.  For the computation of the optical
spectrum, it is enough to consider $A^{\mu}=a$ and~$O=\ud{a}$ in
eqn~(\ref{eq:ThuApr3174049CEST2008}). We obtain the equation 
\begin{equation}
  \label{eq:M2d}
 \partial_\tau\mathbf{v}(t,t+\tau)=\mathbf{M}_1\mathbf{v}(t,t+\tau)
\end{equation}
for the correlators
\begin{equation}
  \label{eq:M2c}
  \mathbf{v}(t,t+\tau)=
  \begin{pmatrix}
    \langle\ud{a}(t){a}(t+\tau)\rangle\\
    \langle\ud{a}(t)b(t+\tau)\rangle
  \end{pmatrix}
\end{equation}
where
\begin{equation}
  \label{eq:M4}
  \mathbf{M}_1=
  \begin{pmatrix}
    M_{\substack{10\\10}} & M_{\substack{10\\01}}\\
    M_{\substack{01\\10}} & M_{\substack{01\\01}}
  \end{pmatrix}
 =
  \begin{pmatrix}
    -i\omega_a-\frac{\Gamma_a}{2} & -ig \\
    -ig & -i\omega_b-\frac{\Gamma_b}{2}
  \end{pmatrix}\,.
\end{equation}

The formal solution follows straightforwardly
from~$\mathbf{v}(t,t+\tau)=e^{\mathbf{M}_1\tau}\mathbf{v}(t,t)$.
Made explicit, it reads (at positive~$\tau$):
\begin{multline}
  \label{eq:M5b}
  \langle\ud{a}(t){a}(t+\tau)\rangle=\frac{\exp[-(\Gamma_{+}-i\frac{\Delta}{2})\tau]}{2 R}\\
  \times\Big\{ e^{i R\tau} [ (R+i\Gamma_--{\Delta}/{2}) n_a(t) -g\, n_{ab}(t)]\\
  {}+e^{-i R\tau}[(R-i\Gamma_-+{\Delta}/{2}) n_a(t) + g\, n_{ab}(t)]\Big\}
\end{multline}
in terms of the \emph{complex Rabi frequency}:
\begin{equation}
  \label{eq:ThuApr3185957CEST2008}
  R=\sqrt{g^2-\left(\Gamma_-+i\frac{\Delta}{2}\right)^2}\,,
\end{equation}
that arises as a direct extension of the dissipationless case,
eqn~(\ref{eq:ThuJun5115514BST2008}). For our discussion, it is
convenient to decompose~$R$ into its real and imaginary parts,
$R=R_\mathrm{r}+iR_\mathrm{i}$.  Out of resonance, the Rabi frequency
is a complex number with both nonzero real and imaginary parts. On the
other hand, at resonance, it is either pure imaginary (in the WC
regime), either pure real (in the SC one). For this latter case it is
worth defining a new quantity:
\begin{equation}
  \label{eq:FriApr4013751UTC2008}
  R_0=R(\Delta=0)=\sqrt{g^2-\Gamma_-^2}\,.
\end{equation}

\begin{figure}[hbpt]
  \centering
  \includegraphics[width=0.45\linewidth]{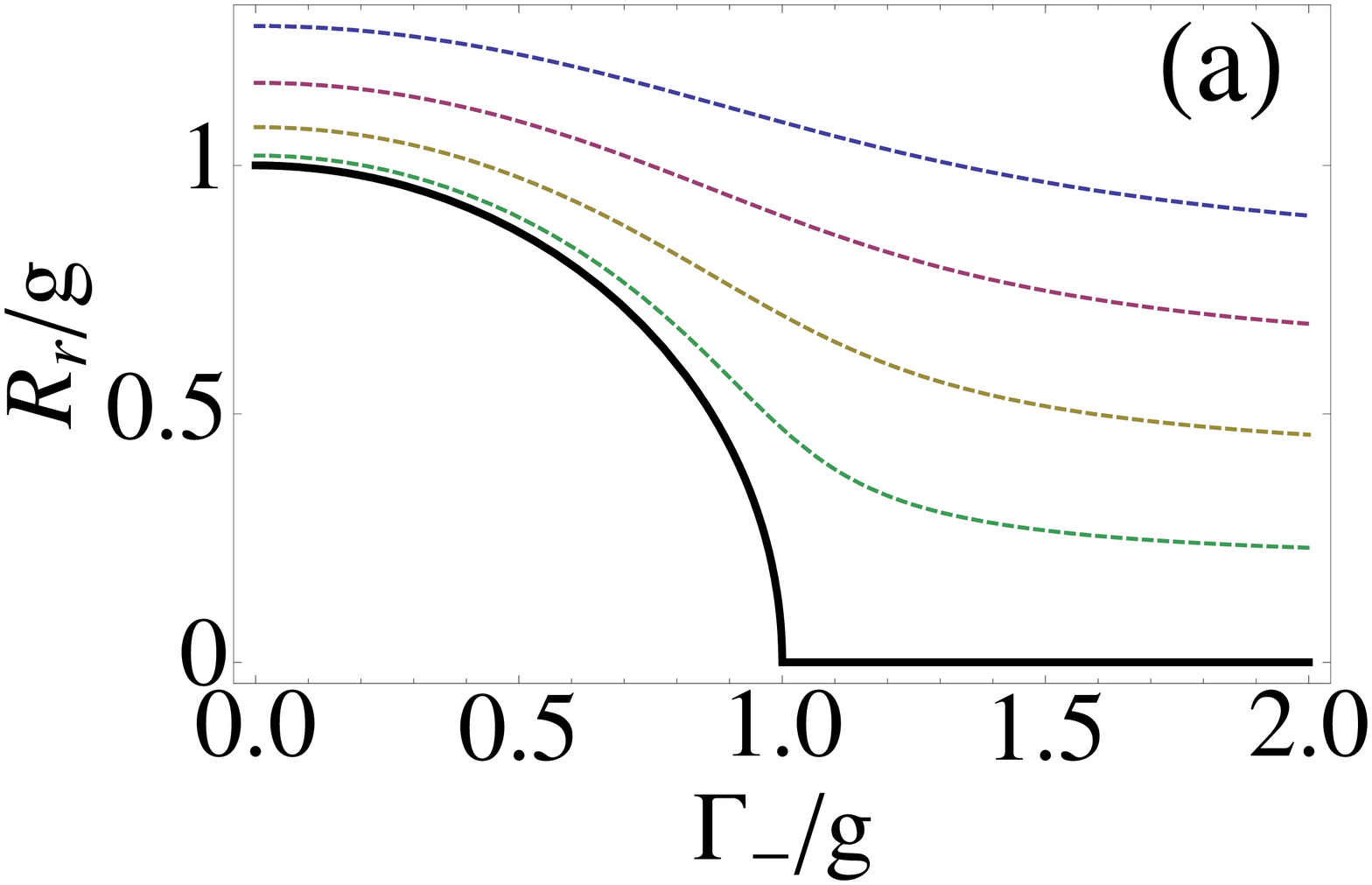}
  \includegraphics[width=0.45\linewidth]{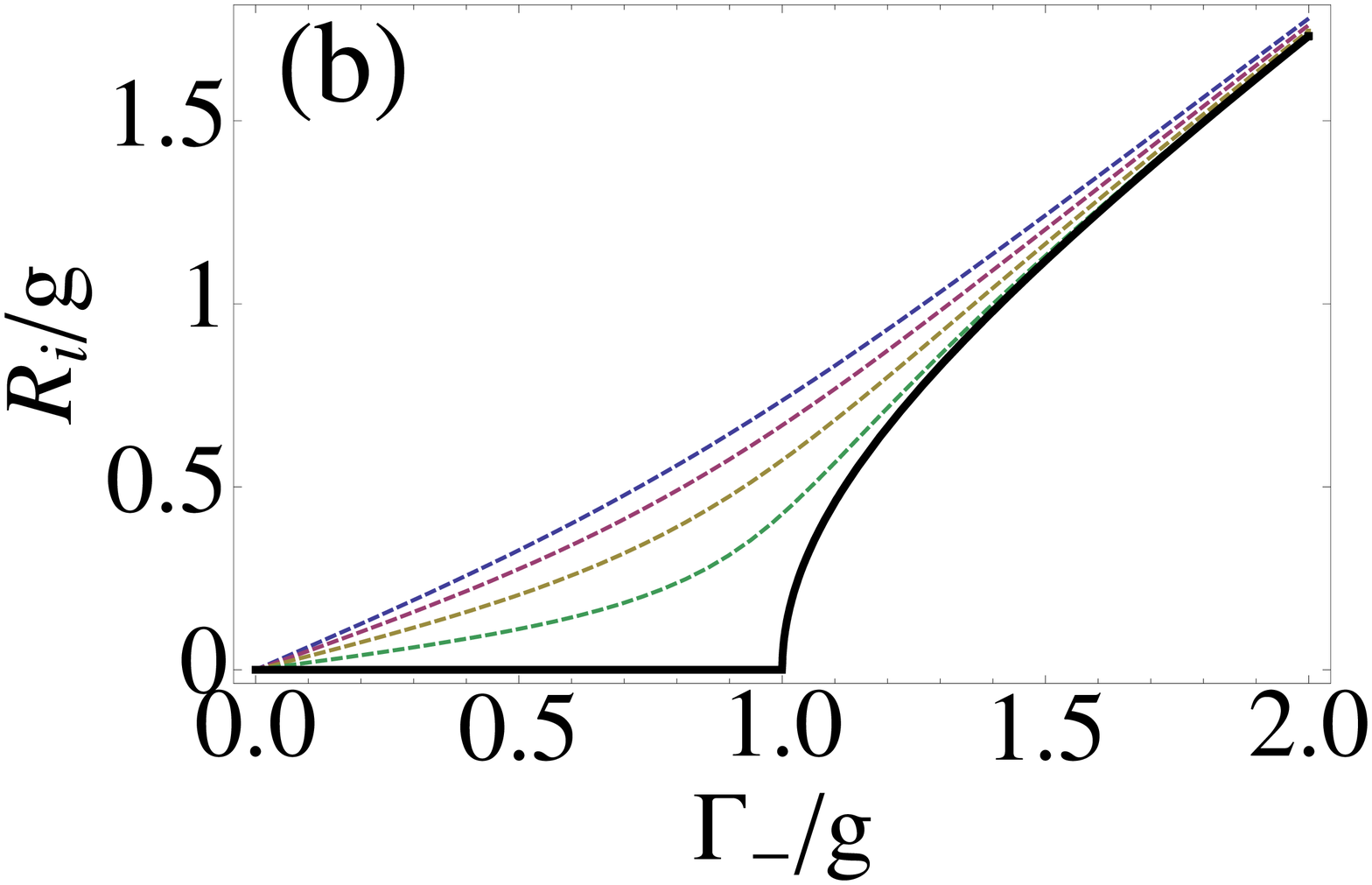}
  \caption{(Color online) Complex Rabi $R/g$, separated in its real
    (a) and imaginary (b) parts, as a function of the decoherence
    parameter $\Gamma_-/g$ for various detunings ($\Delta/g$
    from~$-1.6$, up, to~0, bottom, by steps of~0.4). Solid black lines
    correspond to resonance.}
  \label{fig:Rabi}
\end{figure}

In Fig.~\ref{fig:Rabi}, the real and imaginary parts of $R$ are
plotted as a function of $\Gamma_-/g$ for various detunings.
In the limit of high detuning, $|\Delta|\gg g$, regardless of WC or
SC, the real part becomes independent of the dissipation (decay and
pumping), $R_\mathrm{r}\approx|\Delta|/2$, and the imaginary part
becomes $R_\mathrm{i}\approx\mp\Gamma_{-}$. Therefore, as already
pointed out, the limit of bare modes at energies $\omega_a=0$ and
$\omega_b=-\Delta$ broadened with the bare parameters~$\Gamma_{a/b}$
(FWHM), is always recovered at large detunings.

Using the result of eqn~(\ref{eq:M5b}) into the definition of
eqn~(\ref{eq:ThuApr3172523CEST2008}), we obtain the formal structure
of the emission spectrum:
\begin{equation}
  \label{eq:spe1}
  S(\omega)=\frac{1}{2\pi}\Big[(\mathcal{L}^1+\mathcal{L}^2)-\Im\{\mathcal{C}\}(\mathcal{L}^1-\mathcal{L}^2)-\Re\{\mathcal{C}\}(\mathcal{A}^1-\mathcal{A}^2)\Big]
\end{equation}
with~$\mathcal{L}(\omega)$ and~$\mathcal{A}(\omega)$ some Lorentzian
and dispersive functions whose features (position and broadening) are
entirely specified by the complex Rabi frequency
[eqn~(\ref{eq:ThuApr3185957CEST2008})], $\Gamma_+$
[eqn~(\ref{eq:M2})] and the detuning~$\Delta$:
\begin{subequations}
  \label{eq:parts}
  \begin{align}
    \mathcal{L}^{\substack{1,2}}(\omega)=&\frac{{\Gamma_+\pm
        R_\mathrm{i}}}{({\Gamma_+\pm R_\mathrm{i}})^2+(\omega-({-\frac{\Delta}{2}\mp R_\mathrm{r}}))^2}\,,\\
    \mathcal{A}^{\substack{1,2}}(\omega)=&\frac{\omega-({-\frac{\Delta}{2}\mp
        R_\mathrm{r}})}{({\Gamma_+\pm
        R_\mathrm{i}})^2+(\omega-({-\frac{\Delta}{2}\mp
        R_\mathrm{r}}))^2}\,.
  \end{align}
\end{subequations}
We also introduced~$\mathcal{C}$, a complex coefficient given by
\begin{align}
  \label{eq:spe1cte}
  \mathcal{C}&=\frac{\Gamma_-+i(\frac{\Delta}2+gD)}{R}\,,
\end{align}
that we define in terms of still another parameter, $D$:
\begin{equation}
  \label{eq:D}
  D=\frac{\displaystyle\int_0^\infty\langle\ud{a}b\rangle(t)\,dt}{\displaystyle\int_0^\infty\langle\ud{a}a\rangle(t)\,dt}\,.
\end{equation}

Written in this form, eqns~(\ref{eq:spe1}--\ref{eq:D}) assume a
transparent physical meaning with a clear origin for each term. The
spectrum consists of two peaks (that we label~1 and~2), as is well
known qualitatively for the SC regime. These are composed of a
Lorentzian $\mathcal{L}$ and a dispersive $\mathcal{A}$ part. The
Lorentzian is the fundamental lineshape for a system with a lifetime,
and in the expression above, it inherits most of how the dissipation
gets distributed in the coupled system, including the so-called
\emph{subnatural linewidth averaging} that sees the broadening at
resonance below the cavity mode width~\cite{carmichael89a}. The
dispersive part originates from the coupling as in the Lorentz
(driven) oscillator. In our system, it stems from the driving of one
mode by the other, because of the coupling. This decomposition of each
peak in such terms is therefore entirely clear and expected. More
quantitatively, the first peak, (e.g.,) is centred at
$-\frac{\Delta}{2}-R_\mathrm{r}$ and broadened by
$\Gamma_{+}+R_\mathrm{i}$. As $R_\mathrm{r}>0$, this peak corresponds
to the lower branch ``L''.

So far, all the results hold for both cases of SE and SS. This shows
that the qualitative depiction of SC is robust. This made it possible
to pursue it in a given experimental system with the parameters of the
theoretical models fit for another. This has indeed been the situation
with semiconductor results explained in terms of the formalism built
for atomic systems.

To be complete, the solution now only requires the boundary conditions
that are given by the quantum state of the system. They will affect
the parameter~$D$, eqn~(\ref{eq:D}), that is therefore the bridging
parameter between the two cases. In the next two sections, we address
the two cases and their specificities.

\subsection{Case of Spontaneous Emission}

In the case of Spontaneous Emission~\cite{carmichael89a,andreani99a},
where the system decays from an initial state, the boundary conditions
are supplied for~$\tau=0$ by the initial values $\mathbf{v}(t,t)$,
i.e., the cavity population, $n_a(t)=\langle\ud{a}a\rangle(t)$ and the
coherence element~$n_{ab}(t)=\langle\ud{a}b\rangle(t)$. In turn, those
are completely defined by the initial conditions,
eqns~(\ref{eq:FriApr4021006UTC2008}). Although the analytical
expression for these mean values as a function of time are
cumbersome---see Appendix~\ref{MonMay19214851CEST2008}---the~$D$
coefficient, eqn~(\ref{eq:D}), that determines quantitatively the
lineshape, assumes a (relatively) simpler expression:
\begin{equation}
  \label{eq:Dcoh}
  D^\mathrm{SE}=\frac{[\frac{g}{2}(\gamma_an^0_b-\gamma_bn^0_a)-2in_{ab}^0(\gamma_+^2-\gamma_-^2)](i\gamma_+-\frac{\Delta}2)+2g^2\gamma_+\Re n_{ab}^0 }
  {g^2\gamma_+(n_a^0+n_b^0)+n_a^0\gamma_b(\gamma_+^2+(\frac{\Delta}2)^2){+g\gamma_b(\frac{\Delta}2\Re
      n_{ab}^0+\gamma_+\Im n_{ab}^0)}}\,.
\end{equation}

To prepare the analogy with the SS case in the next section, we also
write the particular case when $n_{ab}^0=0$:
\begin{align}
  \label{eq:Dcoh-nab0}
  D^\mathrm{SE}=\frac{\frac{g}{2}(\gamma_an^0_b-\gamma_bn^0_a)(i\gamma_+-\frac{\Delta}2)}{g^2\gamma_+(n_a^0+n_b^0)+n_a^0\gamma_b(\gamma_+^2+(\frac{\Delta}2)^2)}\,.
\end{align}
This is an important case as it is realized whenever the initial
population of one of the modes is zero. Note that in this case,
$D^\mathrm{SE}$, and therefore also the normalised spectra, do not
depend on the magnitude of the populations.

\subsection{Case of continuous, incoherent pumping}

In the case where the system is excited by a continuous, incoherent
pumping, a steady state is reached and the boundary conditions are
given by the stationary limit, as time tends to infinity, of the
dynamical equation (whose solution is unique). The $D$ parameter,
eqn~(\ref{eq:D}), is defined in this case as:
\begin{equation}
  \label{eq:M3b}
  D^\mathrm{SS}=\frac{n_{ab}^\mathrm{SS}}{n_a^\mathrm{SS}}=
  \frac{\frac{g}2(\gamma_aP_b-\gamma_bP_a)(i\Gamma_+-\frac{\Delta}2)}
  {g^2\Gamma_+(P_a+P_b)+P_a\Gamma_b(\Gamma_+^2+(\frac{\Delta}{2})^2)}\,.
\end{equation}

There is a clear analogy between eqn~(\ref{eq:M3b})---that corresponds
to the SS---and eqn~(\ref{eq:Dcoh-nab0})---that corresponds to SE when
$n_{ab}^0=0$. It is made more meaningful by defining the ratio
\begin{subequations}
  \label{eq:TueMay27011014CEST2008}
  \begin{align}
    \alpha&=n_a^0/n_b^0\qquad(\text{SE case with~$n_{ab}^0=0$})\,,\\
    \alpha&=P_a/P_b\qquad(\text{SS case})\,,
  \end{align}
\end{subequations}
in which case eqns~(\ref{eq:Dcoh-nab0}) and~(\ref{eq:M3b}) assume the
same expression, keeping in mind the definition of
eqns~(\ref{eq:ThuApr3180040CEST2008}). Table~\ref{tab:TueMay27004548CEST2008}
displays this common expression for $D$ in terms of $\alpha$. The
limiting cases when~$\alpha\rightarrow0$ or~$\infty$ are also
given. They correspond to only photons or excitons as the initial
state for the SE, or to the presence of only one kind of incoherent
pumping for the SS case.

The analogy and differences between $D^\mathrm{SE}$ and
$D^\mathrm{SS}$ reflect in the spectra $S^\mathrm{SE}$ and
$S^\mathrm{SS}$. For the same $\alpha$, they become identical when the
pumping rates are negligible as compared to the decays,
$P_{a,b}\ll\gamma_{a,b}$. In this case, where
$\Gamma_{\pm,a,b}\approx\gamma_{\pm,a,b}$, the SS system indeed
behaves like that of the SE of particles that decay independently and
that are, at each emission, either a photon or an exciton, with
probabilities in the ratio~$\alpha$.

However, in the most general case, $D^\mathrm{SS}$ depends on more
parameters---namely, $\alpha$ and $\Gamma_{\pm,a,b}$---than
$D^\mathrm{SE}$, that depends only on $\alpha$ and $\gamma_{\pm,a,b}$,
cf.~eqn~(\ref{eq:ThuApr3180040CEST2008}). Moreover, the pumping rates
$P_{a,b}$ affect $S^\mathrm{SS}$ not only through $\alpha$ and
$D^\mathrm{SS}$, but also in the position and broadening of the peaks
(given by $\Gamma_{\pm}$ and $R$). Therefore, the SS is a more general
case, from which the SE with~$n_{ab}^0=0$ can be obtained, but not the
other way around. On the other hand, as seen in
Table~\ref{tab:TueMay27004548CEST2008}, the SS case cannot recover the
SE case when~$n_{ab}^0\neq0$ (this could be overcome if cross Lindblad
pumping terms were introduced in eqn~(\ref{eq:L2}) with parameters
$P_{ab}$ corresponding to the cross initial mean value $n_{ab}^0$, as
is done in, e.g., Ref.~\cite{delvalle07b}, but this describes another
system).

In any case, an important fact for the semiconductor community is that
a SS with non-vanishing pumping rates is out of reach of the SE of any
initial state, which has been the case studied in the literature so
far~\cite{carmichael89a,andreani99a}, and that even in this limiting
case, the effective quantum state obtained in the SS should still be
resolved self-consistently, rather than assuming for~$\alpha$ the
particular case~$0$ or~$\infty$.

\begin{table}[ht]
  \centering
  \begin{tabular}{| c | c |}
    \hline
    $\displaystyle\alpha=\frac{n_a^0}{n_b^0}=\frac{P_a}{P_b}$ & $D$ \\
    \hline
    \hline
    0 & $\displaystyle\frac{-\frac{g}2(i\Gamma_+-\frac{\Delta}2)\gamma_b}{g^2\Gamma_++\Gamma_b(\Gamma_+^2+(\frac{\Delta}{2})^2))}$\\[1.25em]
    $0<\alpha<\infty$ & $\displaystyle\frac{\frac{g}2(i\Gamma_+-\frac{\Delta}2)(\gamma_a-\gamma_b\alpha)}{g^2\Gamma_+(1+\alpha)+\alpha\Gamma_b(\Gamma_+^2+(\frac{\Delta}{2})^2))}$ \\[1.25em]
    $\infty$ & $\displaystyle\frac{(i\Gamma_{+}-\frac{\Delta}{2})\gamma_a}{2g\Gamma_{+}}$\\
    \hline
  \end{tabular}
  \caption{Expression of~$D$, eqn~(\ref{eq:D}), as a function of~$\alpha$, eqns~(\ref{eq:TueMay27011014CEST2008}), in the SE (with~$\Gamma_{\pm,a,b}\rightarrow\gamma_{\pm,a,b}$ and~$n_{ab}^0=0$) and SS cases. $D$ embodies in the luminescence spectrum the influence of the quantum state of the system. The latter is specified by the initial condition in SE, or the pumping/decay interplay in the SS.}
\label{tab:TueMay27004548CEST2008}
\end{table}

\subsection{Discussion}
\label{WedMay28111634CEST2008}

\begin{figure}[hbpt]
  \centering
  \includegraphics[width=0.5\linewidth]{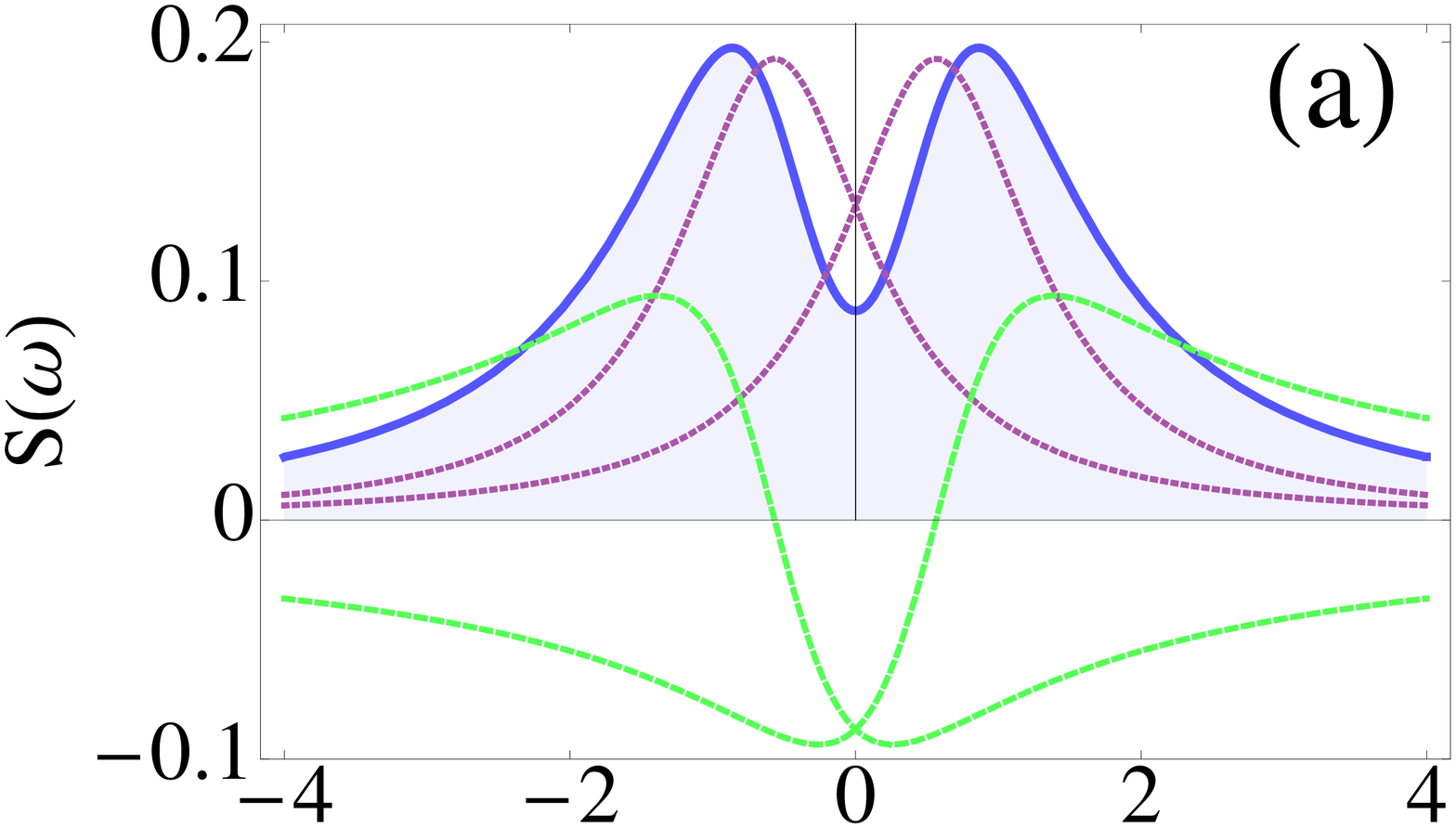}\\
  \includegraphics[width=0.5\linewidth]{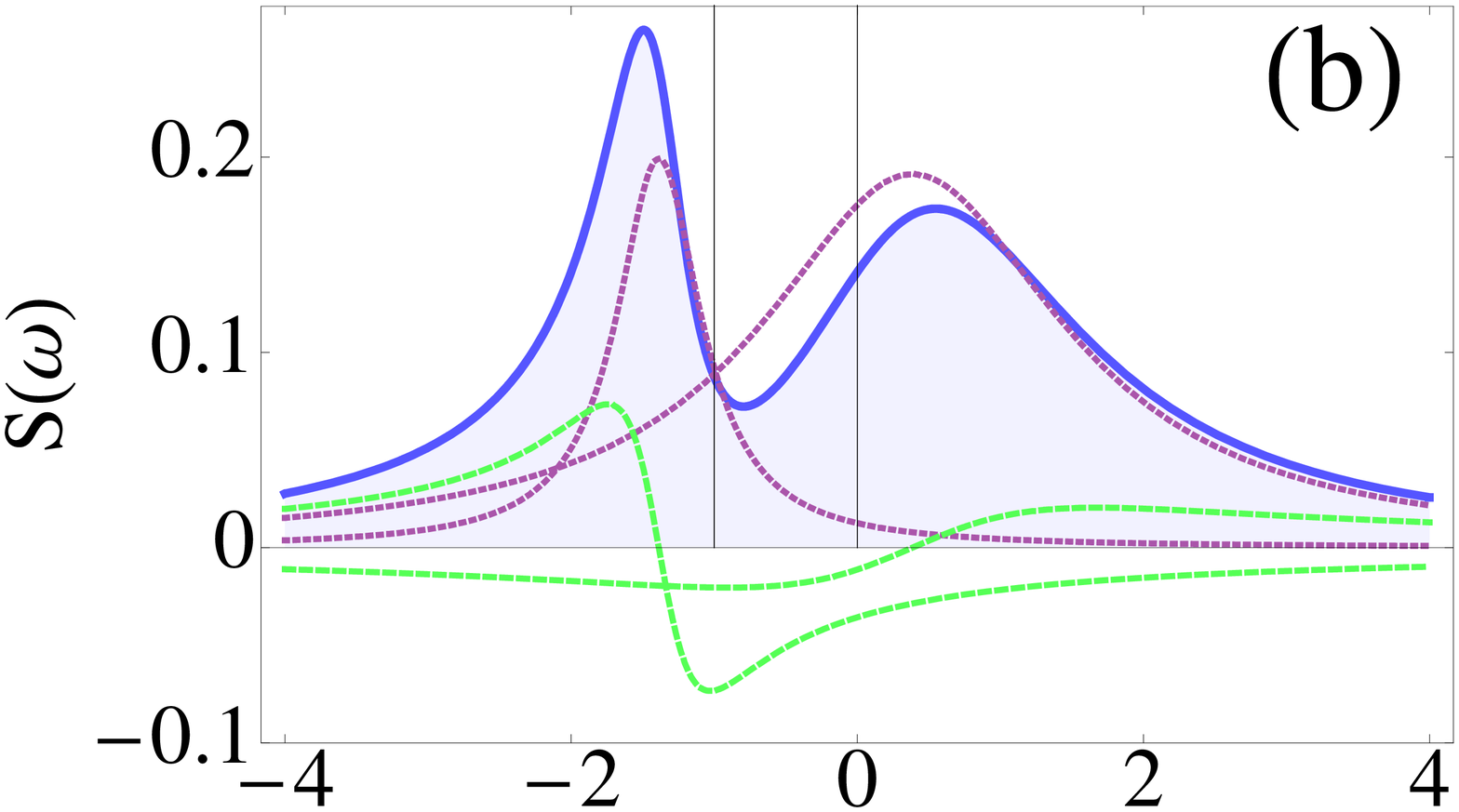}\\
  \includegraphics[width=0.5\linewidth]{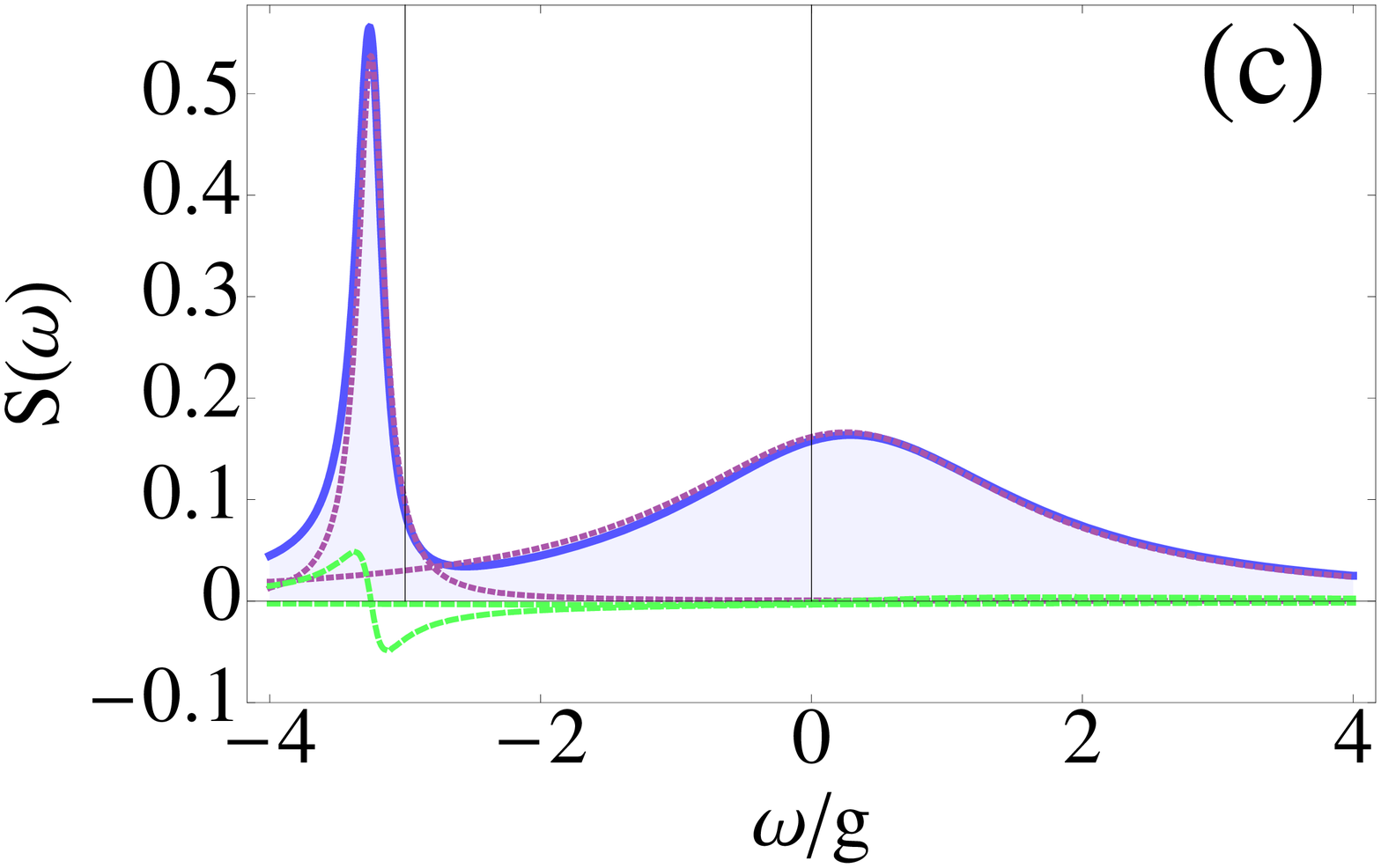}
  \caption{(Color online) Strong-coupling SS spectra (blue solid line)
    and their decomposition into Lorentzian (dotted purple) and
    dispersive parts (dashed green) for various detunings
    ($\Delta/g=0,1,3$) with parameters of point~(c) of
    Fig.~\ref{fig:FriMay16003338UTC2008}: $\gamma_a=3.8g$,
    $\gamma_b=0.1g$, $P_a=0.5g$, $P_b=0.1g$. The vertical black lines
    mark the positions of the bare modes, showing the ``level
    repulsion'' of~SC.}
  \label{fig:spectra}
\end{figure}

With this exposition of the analytical expressions of the luminescence
spectra, and the discussion of their similarity and distinctions that
we have just given, the coverage of the problem is complete and
comprehensive. In order to give a more physical picture of these
abstract results, we shall in the rest of this paper illustrate the
implications that this bears in practical terms. We consider the case
of resonance for this purpose, for reasons detailed in the next
section. This will also allow us to provide self-contained expressions
for the spectra, that can be used more conveniently to plot the
expression or use the model in a nonlinear fitting of the experimental
data~\cite{reithmaier04a, yoshie04a, peter05a}. The system lends
itself naturally to a global fitting, i.e., constraining all fitting
parameters over the various detuning cases, where only the detuning is
allowed to vary, while optimising them
globally~\cite{psip_laussy08a}. There is no difficulty in extending
all the discussions that follow to arbitrary detunings. For instance,
Fig.~\ref{fig:spectra} shows the SS spectra and their mathematical
decompositions into Lorentzian and dispersive parts, as detuning is
varied. Figs.~(b) and~(c) are obtained using
eqn~(\ref{eq:spe1}--\ref{eq:D}), and in this particular case, the
expression~(\ref{eq:M3b}) for~$D$. Fig.~(a), the case at resonance,
can be obtained in the same way, but in next Section we shall bring
all these results together in a condensed expression
[eqn~(\ref{eq:spe5})].

More importantly, as we shall soon appreciate, the resonant case is
the pillar of the SC physics. The main output of the out-of-resonance
case is to help identify or to characterise the resonance, for
instance by localising it in an anticrossing or by providing useful
additional constrains with only one more free parameter in a global
fitting. Even a slight detuning brings features of WC into the SC
system and ultimately, when $\Delta\gg g$, the complex Rabi frequency
converges into the same expression for both regimes (as showed in
Fig.~\ref{fig:Rabi}). This is why we now consider the SC problem in
its purest form: when the coupling between the modes is optimum.

\section{Strong \& Weak Coupling at resonance}
\label{sec:ThuApr3161508CEST2008}

Strong-coupling is most marked at resonance, and this is where its
signature is experimentally ascertained, in the form of an
anticrossing.  Fundamentally, there is another reason why resonance
stands out as predominant: this is where a criterion for SC can be
defined unambiguously in presence of dissipation:\\

\emph{WC and SC are formally defined as the regime where the complex
  Rabi frequency at resonance, eqn~(\ref{eq:FriApr4013751UTC2008}), is
  pure imaginary (WC) or real (SC).}\\

This definition, that takes into account dissipation and pumping,
generalises the classification found in the literature, and is one of
the main results of this work. The reason for this definition is
mainly to be found in the behaviour of the time autocorrelator,
eqn~(\ref{eq:M5b}), that is respectively damped or oscillatory as a
result. The exponential damping is the usual manifestation of
dissipation, that decays the correlations in the field, even when a
steady state is maintained.  On the other hand, in the same situation
of steady averages (no dynamics) but now in SC, the oscillations
with~$\tau$ are the mark of a coherent exchange between the bare
fields (the photon field and exciton field).

\begin{figure}[thbp]
  \centering
  \includegraphics[width=\linewidth]{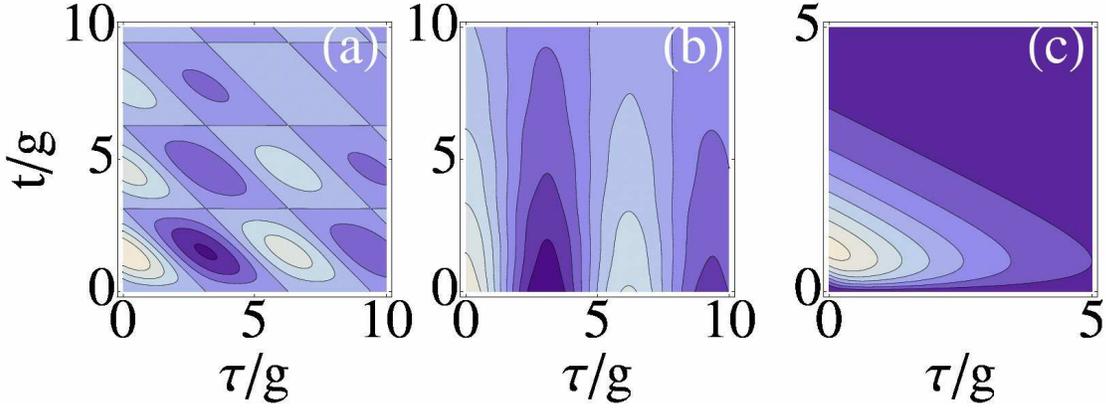}
  \caption{(Color online) Time dynamics of the correlator
    $\Re\langle\ud{a}(t){a}(t+\tau)\rangle^\mathrm{SE}$,
    cf.~eqn~(\ref{eq:M5b}). Only the pattern of oscillations is of
    interest here (lighter blues correspond to higher values). In all
    cases, both the~$t$ and~$\tau$ dynamics tend to zero. Figures (a)
    and~(b) show the SE of an exciton and of an upper polariton,
    respectively, in a very strongly-coupled system ($\gamma_a=0.2g$
    and~$\gamma_b=0.1g$). Fig.~(c) shows the SE of an exciton in
    weak-coupling ($\gamma_a=5.9g$).  The oscillations in~$\tau$,
    rather than in~$t$, are the mark of SC.}
  \label{fig:MonMay19002338UTC2008}
\end{figure}

\begin{figure}[hbpt]
  \centering
  \includegraphics[width=0.45\linewidth]{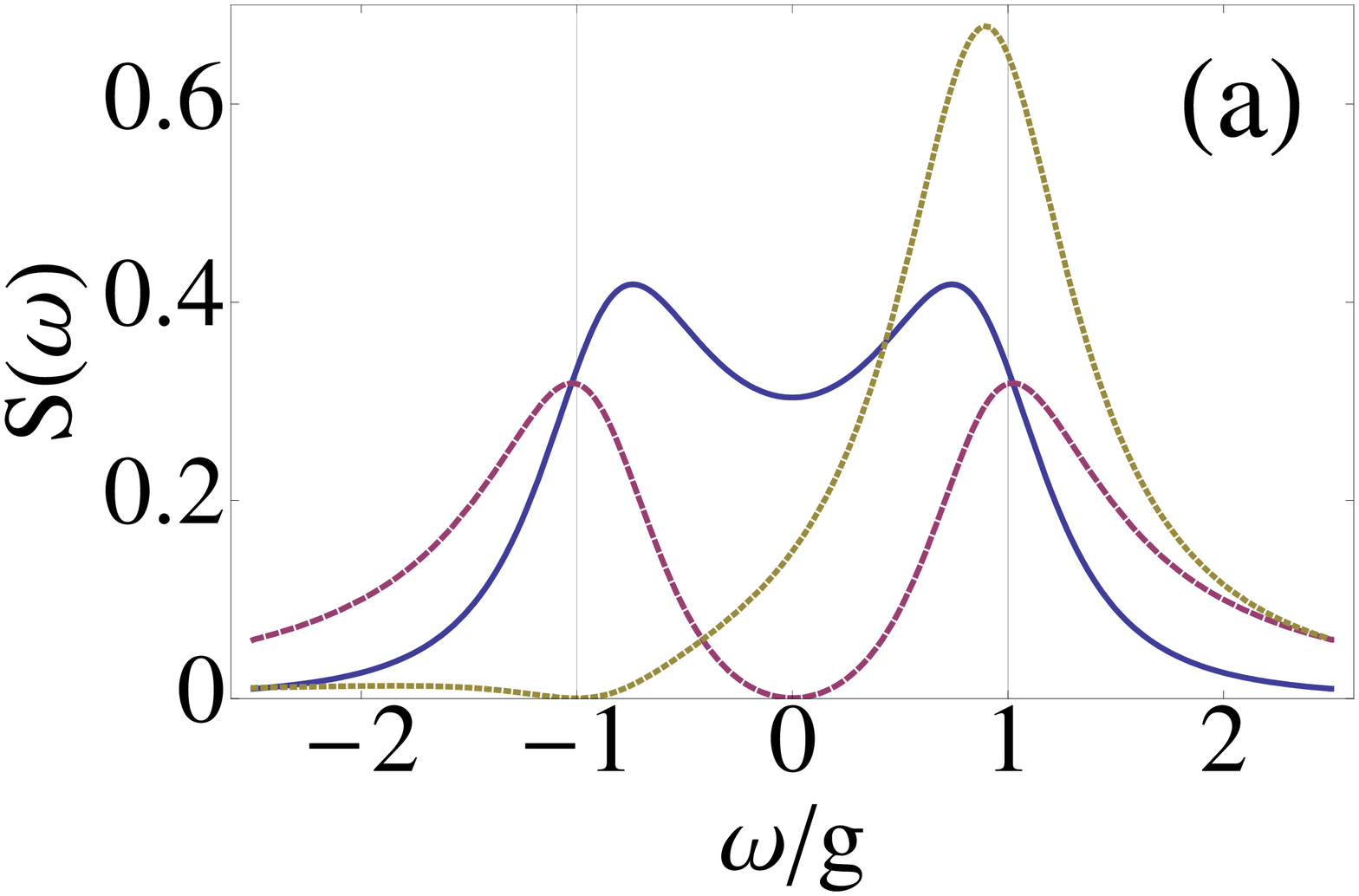}
  \includegraphics[width=0.45\linewidth]{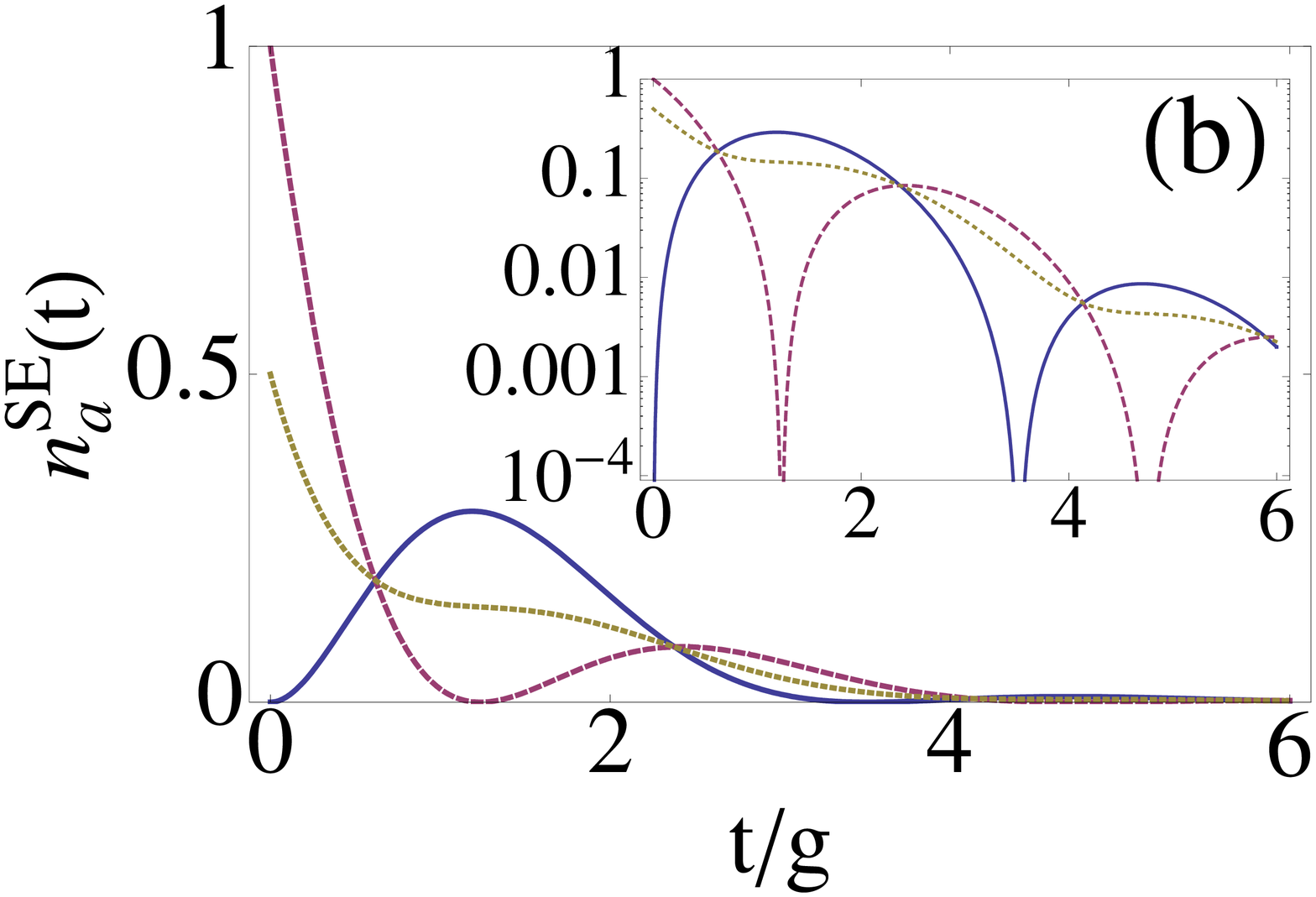}
  \caption{(Color online) Strong-Coupling spectra
    $S_0^\mathrm{SE}(\omega)$ (left) and its corresponding mean number
    dynamics $n_a^\mathrm{SE}(t)$ (right) for the SE of three
    different initial states: In blue solid, one exciton; in purple
    dashed, one photon and in brown dotted, one upper polariton.
    Parameters are $\gamma_a=1.9g$ and $\gamma_b=0.1g$. Inset of (b)
    is the same in log-scale.}
  \label{fig:Spe1}
\end{figure}

\begin{figure}[hbpt]
  \centering
  \includegraphics[width=0.45\linewidth]{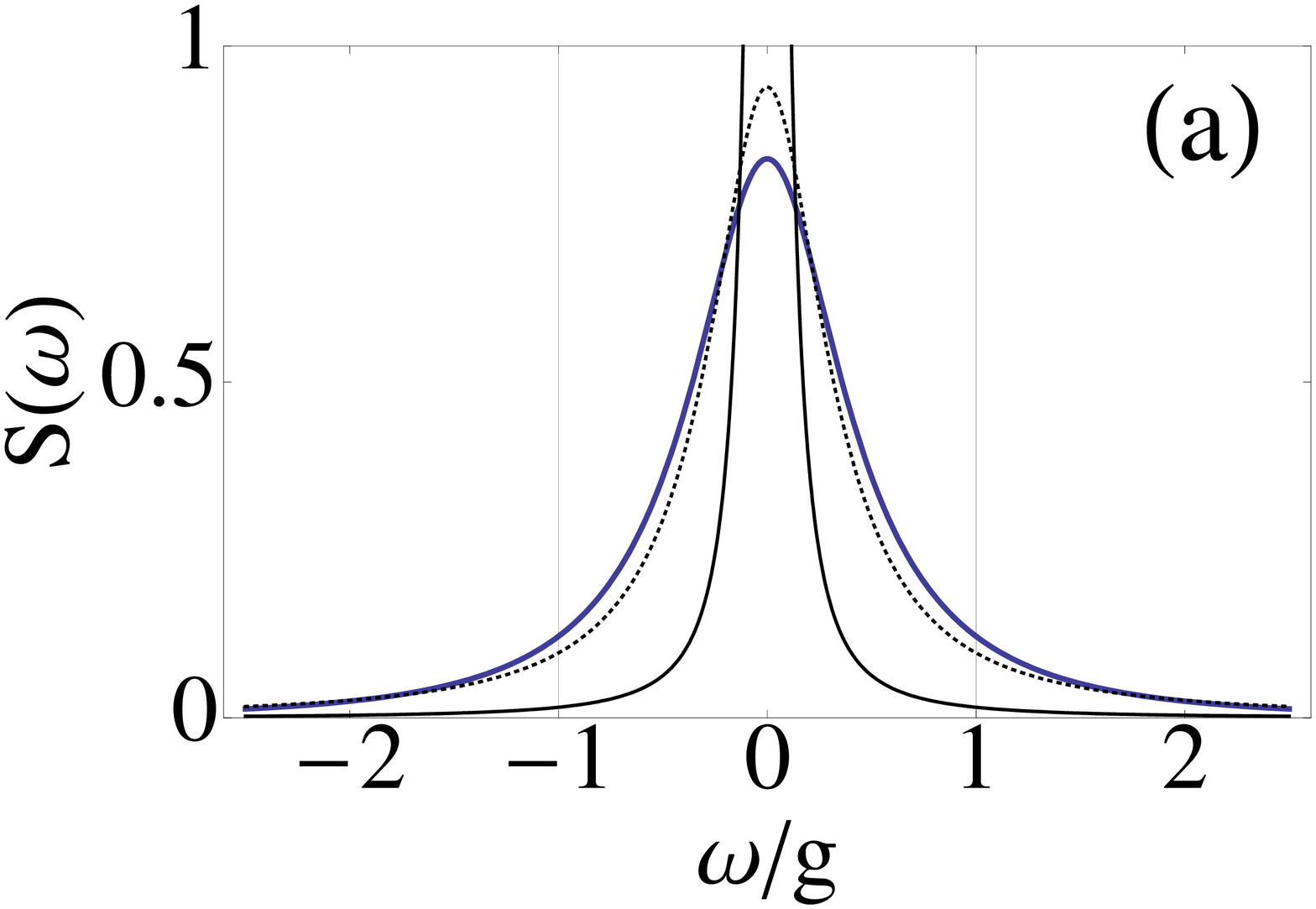}
  \includegraphics[width=0.45\linewidth]{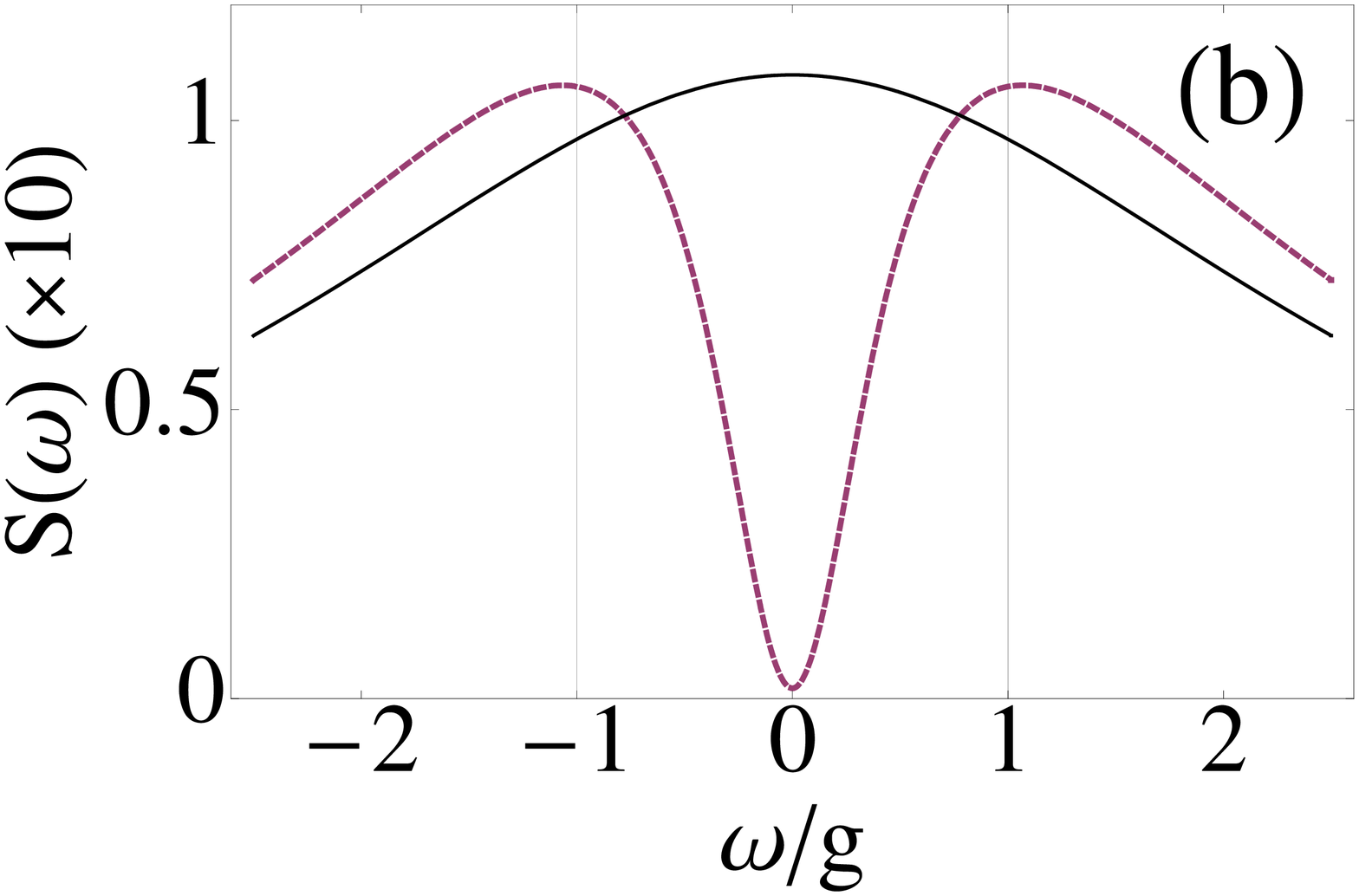}\\
  \includegraphics[width=0.50\linewidth]{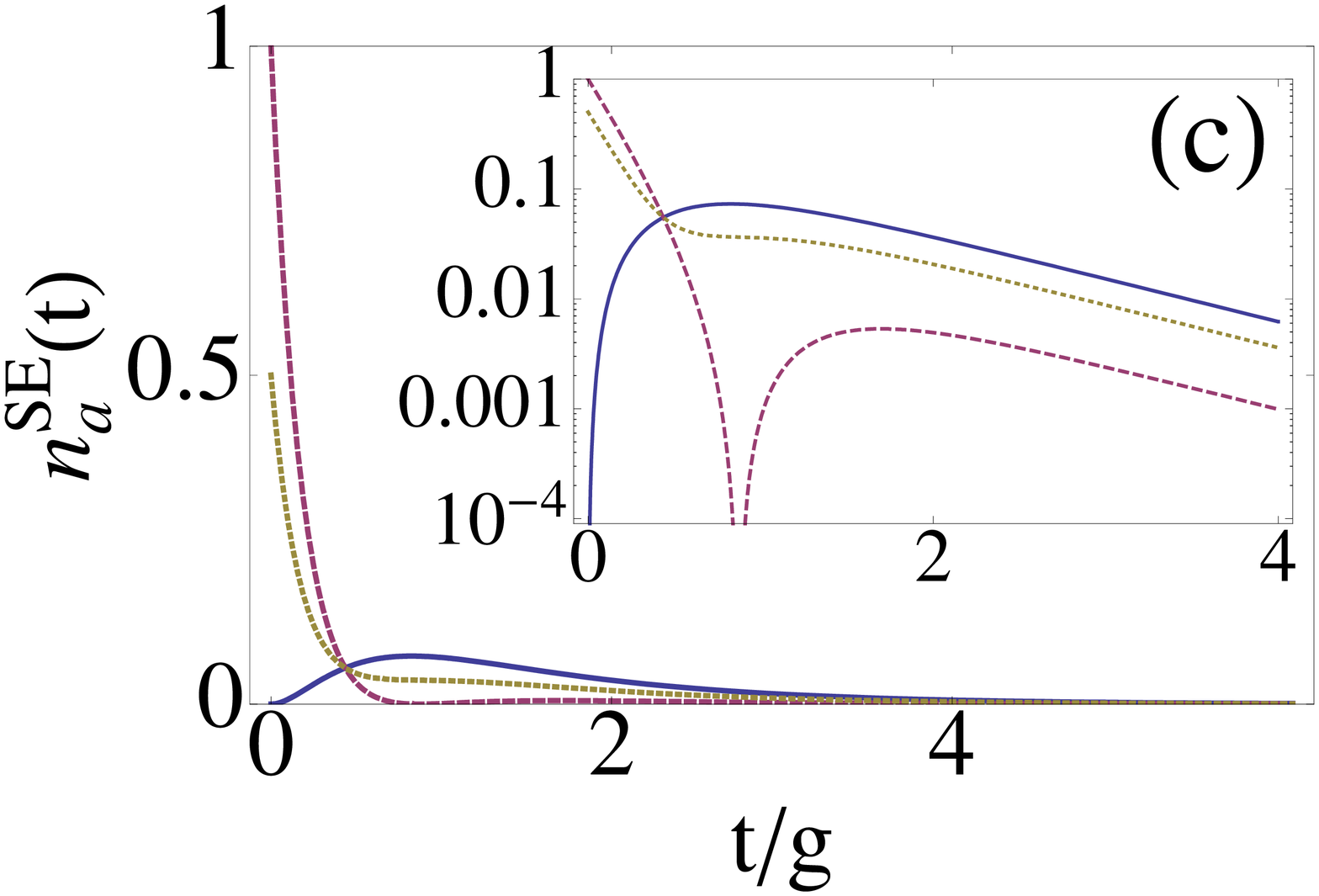}
  \caption{(Color online) Weak-Coupling spectra
    $S_0^\mathrm{SE}(\omega)$ (upper) and its corresponding mean
    number dynamics $n_a^\mathrm{SE}(t)$ (lower) for the SE of an
    exciton (a) and a photon (b) as the initial condition.  In all
    figures, solid blue corresponds to the decay of an exciton; dashed
    purple to the decay of a photon and dotted brown to the decay of
    an upper polariton.  For comparison, we plotted in (a) and (b),
    with solid black lines, the very different bare emission ($g=0$)
    of an exciton and photon respectively. Also in (a), in dashed
    black, that of an exciton decaying with the Purcell rate
    $\gamma_\mathrm{P}=4g^2/\gamma_a$.  Parameters are $\gamma_a=1.9g$
    and $\gamma_b=0.1g$.  Inset of (c) is the same in log-scale.}
  \label{fig:Spe12}
\end{figure}

In the literature, one sometimes encounters the confusion that SC is
linked to a periodic transfer of energy or of population between the
photon and exciton field, or that it follows from a chain of emissions
and absorptions. This is an incorrect general association as one can
explicit cases with apparent oscillations of populations that
correspond to weak coupling, or on the contrary, cases with no
oscillations of populations that are in strong coupling. The two
concepts are therefore not equivalent, as none implies the other. This
is illustrated for the SE case in
Figs.~\ref{fig:MonMay19002338UTC2008}a,~\ref{fig:MonMay19002338UTC2008}b
and~\ref{fig:Spe1} on the one hand, where the system is in SC, and in
Fig.~\ref{fig:MonMay19002338UTC2008}c and~\ref{fig:Spe12} on the other
hand, where it is in WC.  In SS, there is no~$t$ dynamics in any case,
so oscillations of populations are clearly unrelated to weak or strong
coupling. In SE, the distinction is clearly seen in
Fig.~\ref{fig:MonMay19002338UTC2008} where both the~$t$ and~$\tau$
dynamics are shown in a contour-plot in the case where the system is
initially prepared as an exciton, (a) and~(c), or as a polariton,
(b). In the polariton case, the dynamics in~$t$ is simply decaying
(because of the lifetime), while it is clearly oscillating in~$\tau$,
were the proper manifestation of SC is to be found.  The~$t$ decay is
not exactly exponential because in the presence of dissipation, the
polariton is not anymore an ideal eigenstate (the larger the
dissipation, the more the departure). However this effect in SC is so
small that it only consists in a small ``wobbling'' of the~$\tau$
contour lines. On the other hand, the exciton, (a), that is not an
eigenstate, features oscillations both in the~$t$ dynamics (the one
often but unduly regarded as the signature of SC), as well as
the~$\tau$ dynamics. In stark contrast, the exciton in WC, (c),
bounces with~$t$. This, that might appear as an oscillation, is not,
as it happens only once and is damped in the long-time values. This
behaviour is shown quantitatively in Fig.~\ref{fig:Spe1} for SC and
Fig.~\ref{fig:Spe12} for WC, where the population~$n_a(t)$ is
displayed for the SE of an exciton (blue solid), a photon (purple
dashed) and an upper polariton (brown dotted), respectively, along
with the luminescence spectrum that they produce (detected in the
cavity emission). Here it is better seen how, for instance, the
polariton-decay is wobbling as a result of the dissipation, that
perturbs its eigenstate-character and leaks some population to the
lower polariton.  More importantly, note how very different the
spectra are, depending on whether the initial state is a photon or an
exciton, despite the fact that the dynamics is similar in both cases
(see the inset in log-scale of their respective populations). The PL
spectrum observed in the cavity emission is much better resolved when
the system is initially in a photon state, than it is when the system
is initially in an exciton state. The splitting is larger and the
overlap of the peaks smaller in the former case. This will find an
important counterpart in the SS case. In Fig.~\ref{fig:Spe12}, the
corresponding case of WC is shown for clarity, with a decay of
populations and possible oscillations.

\begin{figure}[hbpt]
  \centering
  \includegraphics[width=0.66\linewidth]{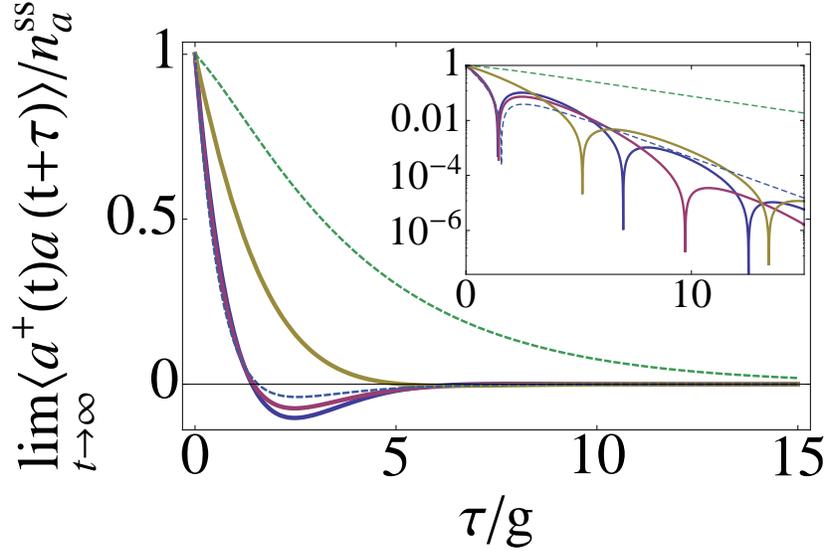}
  \caption{(Color online) Dynamics of
    $\lim_{t\rightarrow\infty}\langle\ud{a}(t){a}(t+\tau)\rangle/n_a^\mathrm{SS}$,
    eqn~(\ref{eq:M5b}) and~(\ref{eq:MonMay19124027CEST2008}), for the
    SS corresponding to the points (a)--(e) in
    Fig.~\ref{fig:FriMay16003338UTC2008}. In inset, the same in
    log-scale.  Solid lines (b, blue; c, purple and d, brown) of SC
    feature oscillations of the correlator, as the mark of SC. Dashed
    lines (a, green and e, blue) correspond to WC. Note that although
    the blue dashed line (e) appears to be similar to other SC lines,
    it does not oscillate in the log-scale, where it only features a
    single local minimum. In the same way, the brown line (d) that
    seems not to oscillate actually features an infinite set of local
    minima, as is revealed in the log scale.}
\label{fig:SunMay18203329UTC2008}
\end{figure}

Figure~\ref{fig:SunMay18203329UTC2008} shows the~$\tau$ dynamics in
the SS (when the~$t$ dynamics has converged and is steady), for five
cases of interest to be discussed later (in
Fig.~\ref{fig:FriMay16003338UTC2008}).  A first look at the dynamics
would seem to gather together a group of two curves that decay
exponentially to good approximation (and remain positive as a result),
and another group of three that assume a local minimum. The correct
classification is the most counter-intuitive in this regard, as it
puts together the dashed lines on the one hand and the solid on the
other, i.e., scrambling them together. The mathematical reason for
this classification is revealed in the inset, where the same dynamics
is plotted on log-scale. The dashed (resp.~solid) lines correspond to
parameters where the system is in WC (resp.~SC) according to our
definition, i.e., to values of~$R$ that are imaginary on the one hand and
real on the other. In log-scale, this corresponds respectively to a
damping of the correlator, against oscillations with an infinite number
of local minima. Note that the blue dashed line features one local
minimum, which does not correspond to an oscillatory---or
coherent-exchange---behaviour of the fields, but rather to a jolt in
the damping. These considerations that may appear abstract at this
level will later turn out to show up as the actual emergence of
splitted (dressed) states or not in the emitted spectrum.

We now return to the general (SE/SS) expression for the spectra,
eqn~(\ref{eq:spe1}), that, at resonance in SC, simplifies to:
\begin{equation}
  \label{eq:spe2}
  S_0(\omega)=\frac{1}{2\pi}\Bigg[(\mathcal{L}_\mathrm{s}^1+\mathcal{L}_\mathrm{s}^2)-\frac{g\Re\{D_0\}}{R_0}(\mathcal{L}_\mathrm{s}^1-\mathcal{L}_\mathrm{s}^2)\\
  +\frac{g\Im\{D_0\}-\Gamma_{-}}{R_0}(\mathcal{A}_\mathrm{s}^1-\mathcal{A}_\mathrm{s}^1)\Bigg]
\end{equation}
where we used the definition for the Rabi frequency at resonance,
eqn~(\ref{eq:FriApr4013751UTC2008}), and
\begin{subequations}
  \label{eq:parts2}
  \begin{align}
    \mathcal{L}_\mathrm{s}^{\substack{1,2}}(\omega)=&\frac{\Gamma_+}{\Gamma_+^2+(\omega\pm R_0)^2}\,,\\
    \mathcal{A}_\mathrm{s}^{\substack{1,2}}(\omega)=&\frac{\omega\pm
      R_0}{\Gamma_+^2+(\omega\pm R_0)^2}\,.
  \end{align}
\end{subequations}

In the weak coupling regime, with $R_0$ pure imaginary ($g<|\Gamma_-|$),
the positions of the two peaks collapse onto the centre,
$\omega_a=\omega_b=0$.  Defining $iR_\mathrm{w}=R_0$, with
$R_\mathrm{w}=\sqrt{\Gamma_-^2-g^2}$ a real number, the general
expression for the spectra rewrites as:
\begin{align}
  \label{eq:spe1weak}
  S_0^\mathrm{w}(\omega)&=\frac{1}{2\pi}\left(1+ \frac{\Gamma_--g\Im\{D_0\}}{R_\mathrm{w}}\right)\mathcal{L}_\mathrm{w}^1\nonumber\\
  &+\frac{1}{2\pi} \left(1-\frac{\Gamma_--g\Im\{D_0\}}{R_\mathrm{w}}\right)  \mathcal{L}_\mathrm{w}^2\nonumber\\
  &-\frac{1}{2\pi}\frac{g\Re\{D_0\}}{R_\mathrm{w}}(\mathcal{A}_\mathrm{w}^1-\mathcal{A}_\mathrm{w}^2)\,,
\end{align}
with the Lorentzian and dispersive contributions now given by:
\begin{subequations}
  \begin{align}
    \label{eq:partWeak}
    \mathcal{L}_\mathrm{w}^{\substack{1,2}}(\omega)=&\frac{{\Gamma_+\pm
        R_\mathrm{w}}}{({\Gamma_+\pm R_\mathrm{w}})^2+\omega^2}\\
    \mathcal{A}_\mathrm{w}^{\substack{1,2}}(\omega)=&\frac{\omega}{(\Gamma_+\pm
      R_\mathrm{w})^2+\omega^2}\,.
  \end{align}
\end{subequations}

We now address the specifics of the SE and SS cases.

\subsection{Case of Spontaneous Emission}

In the most general case of SE, the~$D^\mathrm{SE}$ coefficient at
resonance, $D_0^\mathrm{SE}$, is a complex number.
% %
% \begin{equation}
%   \label{eq:D0coh}
%   D_0^\mathrm{SE}=\frac{i\frac{g}{2}\gamma_+(\gamma_a n^0_b-\gamma_b n^0_a)+2n_{ab}^0\gamma_+(\gamma_+^2-\gamma_-^2)+2g^2\gamma_+ \Re n_{ab}^0}{g^2\gamma_+ (n_a^0+n_b^0)+n_a^0\gamma_b\gamma_+^2+(\Re n_{ab}^0+\gamma_+\Im n_{ab}^0)g\gamma_b}\,.
% \end{equation}
%
If the initial condition further fulfils $\Re n_{ab}^0=0$, it becomes
pure imaginary.  Usually, the initial states considered are
independent states of photons or excitons (not a quantum
superposition), where indeed
$n_{ab}^0=0$~\cite{carmichael89a,andreani99a}. In these cases,
\begin{equation}
  \label{eq:D0coh2}
  D_0^\mathrm{SE}=i\frac{\frac{g}{2}(\gamma_an^0_b-\gamma_bn^0_a)}{g^2(n_a^0+n_b^0)+n_a^0\gamma_b\gamma_+}\,,
\end{equation}
which yields the following expression for the spectrum:
\begin{equation}
  \label{eq:speCoh1}
  S_0^\mathrm{SE}(\omega)=\frac{1}{\pi}
  \frac{\frac{\gamma_a + \gamma_b}{2} (g^2 + \frac{\gamma_a
      \gamma_b}{4}) (g^2 n_b^0 + \frac{n_a^0 \gamma_b^2}{4} + n_a^0)}{\left( \omega^4 +\omega^2(\frac{\gamma_a^2 + \gamma_b^2}{4} - 2 g^2) + (g^2 + \frac{\gamma_a \gamma_b}{4})^2
    \right) \left(g^2 (n_a^0 + n_b^0) + n_a^0 \gamma_b
      \frac{\gamma_a + \gamma_b}{4} \right)}\,.
\end{equation}

As noted earlier, the SE spectrum of exciton observed in the leaky
modes is obtained from eqn~(\ref{eq:speCoh1}) by exchanging the
indices $a\leftrightarrow b$. We illustrate this with the two
particular cases that follow.

The typical detection geometry for the spontaneous emission of an atom
in a cavity consists in having the atom in its excited state as the
initial condition, and observing its direct emission spectrum. In this
case the role of the cavity is merely to affect the dynamics of its
relaxation, that is oscillatory with the light-field in the case of
SC. This case corresponds to~$n^0_{b}=1$ and~$n^0_{a}=n^0_{ab}=0$ in
eqn~(\ref{eq:speCoh1}) with~$a\leftrightarrow b$. This
gives~\cite{carmichael89a}:
\begin{equation}
  \label{eq:speCoh2}
  S_0^\mathrm{SE}(\omega)=\frac{1}{\pi}
  \frac{\frac{\gamma_a + \gamma_b}{2} (g^2 + \frac{\gamma_a\gamma_b}{4}) ( \frac{\gamma_{a}^2}{4}+\omega^2)}
  {\omega^4+\omega^2 (\frac{\gamma_a^2 + \gamma_b^2}{4} - 2 g^2)+(g^2+\gamma_{a} \frac{\gamma_a + \gamma_b}{4})(g^2 + \frac{\gamma_a\gamma_b}{4})^2}\,.
\end{equation}

In the semiconductor case, one would typically still have in mind
the excited state of the exciton as the initial condition, but this
time, this is the cavity emission that is probed. The initial
condition is therefore the same as before but without
interchanging~$a$ and~$b$ in eqn~(\ref{eq:speCoh1}), which reads in
this case:
\begin{equation}
  \label{eq:spe7}
  S_0^\mathrm{SE}(\omega)=\frac{1}{\pi}\frac{2(\gamma_a+\gamma_b)(4g^2+\gamma_a\gamma_b)}
  {16\omega^4-4\omega^2(8g^2-\gamma_a^2-\gamma_b^2)+(4g^2+\gamma_a\gamma_b)^2}\,.
\end{equation}

The difference in the lineshape due to the initial quantum state is
seen in Fig.~\ref{fig:Spe1}. The visibility of the line-splitting is
much reduced in the case of an exciton in SC which SE is detected
through the cavity emission, than in the case of a photon. With a
polariton as an initial state, only one line is produced.

Again, by reason of symmetry, interchanging~$a\leftrightarrow b$ in
eqns~(\ref{eq:speCoh2}) and (\ref{eq:spe7}), correspond to the SE of
the system prepared as a photon at the initial time and detected in,
respectively, the cavity emission on the one hand
(eqn~\ref{eq:speCoh2}, $a\leftrightarrow b$), and in the leaky mode
emission on the other hand (eqn \ref{eq:spe7}). In the latter case,
the spectrum is invariant under the exchange $a\leftrightarrow b$.
Fig.~(\ref{fig:Spe1}) also hints to the changes brought by the
detection channel (direct emission of the exciton or through the
cavity mode).

If~$n^0_a=0$ or~$n^0_b=0$ (in which case~$n^0_{ab}=0$), the normalized
spectra do not depend on the nonzero value $n_b^0$ or $n_a^0$.  That
is, one cannot distinguish in the lineshape, the decay of one exciton
from that of two, or more. In the more general case, when
$n_{ab}^0\neq0$, the peaks can be differently weighted. For instance,
starting with an upper polariton $|\phi_0
\rangle=(|1,0\rangle+|0,1\rangle)/\sqrt{2}$
($n_{a}^0=n_b^0=n_{ab}^0=1/2$) gives rise to a dominant
upper-polariton peak (labelled~2 in the above equations, as seen in
the brown dotted line in Fig.~\ref{fig:Spe1}).  One can classify the
possible lineshapes obtained for various initial states.  For
instance, as we have just mentioned, the normalized spectrum
of~$\ket{0,n}$ as an initial state, is the same whatever the
nonzero~$n$, which is not entirely unexpected from a linear model.
From the previous statement, the same spectrum is also obtained for a
coherent state or a thermal state of photons, or indeed any quantum
state, as long as the exciton population remains zero. In the same
way, the PL spectrum of the product of coherent states in the photon
and exciton fields, $\ket{z}\ket{z'}$ with~$z=z'\in\mathbb{C}^*$, is
the same as that of a polariton state~$\ket{0,1}+\ket{1,0}$, although
both are very different in character: a classical state on the one
hand and a maximally entangled quantum state on the other.

\subsection{Case of continuous, incoherent pumping}

In the SS, at resonance, $D_0^\mathrm{SS}$ is pure imaginary:
\begin{equation}
  \label{eq:D0ss}
  D_0^\mathrm{SS}=i\frac{\frac{g}2(\gamma_aP_b-\gamma_bP_a)}
  {g^2(P_a+P_b)+P_a\Gamma_b\Gamma_+}\,,
\end{equation}
and the term that consists in the difference of Lorentzians in
eqn~(\ref{eq:spe1}) disappears: $\Im\{\mathcal{C}\}=0$. As a result,
the two peaks are equally weighted for any combination of parameters:
\begin{align}
  \label{eq:spe3}
  S_0^\mathrm{SS}(\omega)=&\frac{1}{2\pi}\Bigg[(\mathcal{L}_\mathrm{s}^1+\mathcal{L}_\mathrm{s}^2)+\frac{g
    \Im\{D_0^\mathrm{SS}\}-\Gamma_{-}}{R_0}(\mathcal{A}_\mathrm{s}^1-\mathcal{A}_\mathrm{s}^2)\Bigg]\,.
\end{align}

The only way to weight more one of the peaks than the other in the SS
of an incoherent pumping, would be to pump directly the polariton
(dressed) states, as is the case in higher-dimensional systems were
polaritons states with nonzero momentum relax into the ground
state~\cite{laussy04c} or in 0D case when cross pumping is
considered~\cite{delvalle07b}. In our present model, however, such
terms are excluded. The two peaks of the Rabi doublet, composed of a
Lorentzian and a dispersive part, are both symmetric with respect to
$\omega_a=\omega_b=0$. Only if $\Im\{D_0^\mathrm{SS}\}=\Gamma_{-}/g$,
the spectrum of eqn~(\ref{eq:spe3}) consists exclusively of two
Lorentzians. The parameters that correspond to this case are those
fulfilling either~$g^2=\frac{P_a}{P_b-P_a}\Gamma_b\Gamma_{-}$ or
$\Gamma_{+}=0$. The second case corresponds to the limiting case of
diverging populations, where the SC becomes arbitrarily good. In the
most general case, the dispersive part will contribute to the fine
quantitative structure of the spectrum, bringing closer or further
apart the maxima and thus altering the apparent magnitude of the Rabi
splitting. In some extreme cases, as we shall discuss, it even
contrives to blur the resolution of the two peaks and a single peak
results, even though the modes split in energy. As for the weak
coupling formula, it simplifies to:
\begin{align}
  \label{eq:spe1weakInc}
  S_0^\mathrm{w}(\omega)&=\frac{1}{2\pi}\left(1+ \frac{\Gamma_--g\Im\{D_0\}}{R_\mathrm{w}}\right)\mathcal{L}_\mathrm{w}^1\nonumber\\
  &+\frac{1}{2\pi} \left(1-\frac{\Gamma_--g\Im\{D_0\}}{R_\mathrm{w}}\right)  \mathcal{L}_\mathrm{w}^2\,.
\end{align}
Both decompositions, eqns~(\ref{eq:spe3}) and (\ref{eq:spe1weakInc}),
have been given to spell-out the structure of the spectra in both
regimes. The unified expression that covers them both reads
explicitly:
\begin{equation}
  \label{eq:spe5}
  S_{0}^\mathrm{SS}(\omega)=\frac{1}{\pi n_{a}^\mathrm{SS}}\frac{8g^2P_b+2P_a(4\omega^2+\Gamma_b^2)}
  {16\omega^4-4\omega^2(8g^2-\Gamma_a^2-\Gamma_b^2)+(4g^2+\Gamma_a\Gamma_b)^2}\,.
\end{equation}
It is the counterpart for SS of eqn~(\ref{eq:speCoh1}), for SE. The
case of excitonic emission can also be obtained, as for SE, by
interchanging the indices $a\leftrightarrow b$.

\begin{figure}[h]
  \centering
  \includegraphics[width=.9\linewidth]{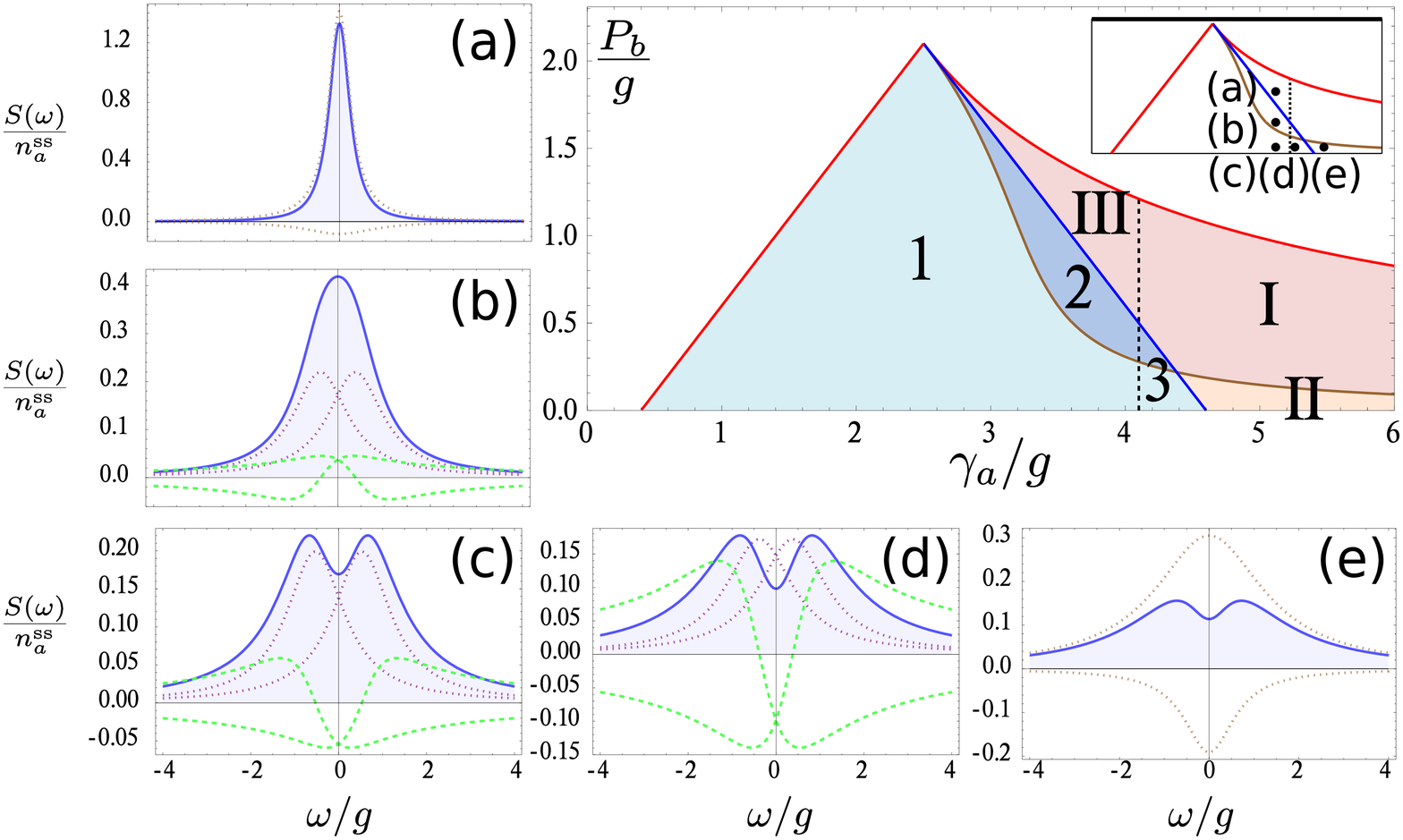}
  \caption{(Color online) Phase space of the SS Strong/Weak Coupling
    as a function of $P_b/g$ and $\gamma_a/g$ for the parameters
    $\gamma_b=0.1g$ and $P_a/g=0.5g$. The red lines,
    eqns~(\ref{eq:MonMay19125701CEST2008}--\ref{eq:MonMay19125724CEST2008}),
    delimit the region where there is a steady state. The blue line,
    eqn~(\ref{eq:MonMay19125840CEST2008}), separates the strong (in
    shades of blue) from the weak (shades of red) coupling regions.
    The dotted black line, $\gamma_a=4g+\gamma_b$, separates SC and WC
    regions in the absence of pumping. The brown line,
    eqn~(\ref{eq:MonMay19130211CEST2008}), separates the regions where
    one (dark blue) or two (light blue) peaks can be resolved in the
    luminescence spectra.  This delineates three areas in the SC
    region: (1) two peaks are resolved in the spectra, (2) the two
    peaks cannot be resolved and effectively merge into one, albeit in
    SC, and (3) SC is achieved thanks to the pump~$P_a$ (with one or
    two peaks visible depending of the overlap with the light or dark
    area) despite the large dissipation that predicts WC according to
    eqn~(\ref{eq:MonMay19131126CEST2008}).  In the same way we can
    distinguish three regions in weak coupling: (I) standard WC, (II)
    SC with a two peaked spectrum and (III) WC due to pumping $P_b$.
    The surrounding figures (a) to (e) show spectra from these regions
    and their decomposition into, Lorentzian and dispersive parts in
    SC, or Lorentzians centred at zero in the WC.  Parameters
    correspond to the points in the inset: (a) $\gamma_a=3.8g$ and
    $P_b=g$, (b) $\gamma_a=3.8g$ and $P_b=0.5g$, (c) $\gamma_a=3.8g$
    and $P_b=0.1g$, (d) $\gamma_a=4.2g$ and $P_b=0.1g$, (e)
    $\gamma_a=4.8g$ and $P_b=0.1g$. Observe how, in SC, two
    eigenstates have emerged in purple dotted lines, even in the
    cases---like in (b)---where they are not seen in the total
    spectrum. In the same way, in WC, all the emission emanates from
    the origin, although a two-peak structure can arise as a result of
    a resonance, also centred at the origin.}
  \label{fig:FriMay16003338UTC2008}
\end{figure}

\subsection{Discussion}
\label{ThuJun5173453BST2008}

Although the spectra in the semiconductor case that are probed at
negligible electronic pumping ($P_b\ll1$) with no cavity pumping at
all ($P_a=0$), are in principle described by the same expression as
that of the SE case used in the atomic model, in practise, however,
both of these conditions can be easily violated. The renormalization
of~$\gamma_b$ with $P_b$ brings significant corrections well into the
linear regime.  For instance, for parameters of point~(c) in
Fig.~\ref{fig:FriMay16003338UTC2008} with~$P_a=0$, the rate~$P_b$ that
is needed to bring a~$100\%$ correction to~$\gamma_b$ yields,
according to eqns~(\ref{eq:MonMay19124027CEST2008})
and~(\ref{eq:SunMay25222529CEST2008}), averages much below unity,
namely, $n_a^\mathrm{SS}\approx0.026$
and~$n_b^\mathrm{SS}\approx0.121$. By the time $n_b$ reaches unity,
with $n_a$ still one fourth smaller, the correction on the effective
decay rate has became 400\%. As this is~$n_a$ which is proportional to
the signal detected in the laboratory, the electronic pumping must be
kept very small so that corrections to the effective linewidth can be
safely neglected. As the more interesting nonlinear regime is probed,
the renormalized $\Gamma$s become very different from the
bare~$\gamma$s.

Second, even in the vanishing electronic pumping limit, it must be
held true that~$P_a$ is zero. Even if only an electronic pumping is
supplied externally by the experiment, the pumping rates of the model
are the effective excitation rates of the cavity and exciton field
inside the cavity, and it is clear that photons get injected in the
cavity in structures that consists of numerous spectator dots
surrounding the one in SC
(cf.~Fig.~\ref{fig:FriMar14002142UTC2008}). Although most of these
dots are in WC with the cavity, they affect the dynamics of the SC QD
by pouring cavity photons in the system. In the steady state,
following our previous discussion, this corresponds to changing the
effective quantum state for the emission of the strongly-coupled
QD. As we shall see in more detail in what follows, this bears huge
consequences on the appearance of the emitted doublet, especially on
its visibility.

To fully appreciate the importance and deep consequences of these two
provisions made by the SS case on its SE counterpart, we devote the
rest of this Section to a vivid representation in the space of pumping
and decay rates parameters. Now that it has been made clear what is
the relationship of the SE case with respect to the SS one, we shall
focus on the latter that is the adequate, general formalism to
describe SC of QDs in microcavities.

In presence of a continuous, incoherent pumping, the criterion for
strong-coupling---from the requirement of energy splitting and
oscillations in the~$\tau$ dynamics that we have discussed
above---gets upgraded from its usual expression~\cite{carmichael89a}:
\begin{equation}
  \label{eq:MonMay19131126CEST2008}
  g>|\gamma_-|\,,
\end{equation}
to the more general condition:
\begin{equation}
  \label{eq:MonNov5033739UTC2007}
  g>|\Gamma_-|\,.
\end{equation}

The quantitative and qualitative implications and their extent are
shown in Fig.~\ref{fig:FriMay16003338UTC2008}, where we have fixed the
parameters~$\gamma_b=0.1g$ and $P_a/g=0.5g$, and outlined the various
regions of interest as~$P_b$ and~$\gamma_a$ are varied. This choice of
representation allows us to investigate the changes that can be easily
imprinted experimentally in the system: by tuning~$P_b$ on a system of
varying quality with regard to SC, as specified by its quality factor
that tunes~$\gamma_a$.

The red lines, that enclose the filled regions, delimit a frontier
above which the pump is so high that populations diverge (there is no
steady state).  This is expressed mathematically by the condition that
the common denominator of the mean values in
eqns~(\ref{eq:WedMay28004656CEST2008}) vanishes.  It is equivalent to
setting the total decay rate of the system---that should remain
positive in order for the correlators to decay in time---to zero.  In
the SC regime, such a rate is given by $\Gamma_+$ and therefore the
condition for a SS reads:
\begin{equation}
  \label{eq:MonMay19125701CEST2008}
  P_b=\gamma_a+\gamma_b-P_a\,,\quad(\mathrm{i.e.,~}\Gamma_+=0)\,.
\end{equation}
In the WC regime, the system decays with an effective Purcell rate and
the condition transforms into
\begin{equation}
  \label{eq:MonMay19125724CEST2008}
  P_b=\frac{4g^2}{\gamma_a+\gamma_b-P_a}\,.
\end{equation}

The main separation inside that region where a SS exists, is that
between SC (in shades of blue, inside the triangle) and WC (in shades
of red, on its right elbow).  The blue solid line that marks this
boundary, specified by
\begin{equation}
  \label{eq:MonMay19125840CEST2008}
  P_a=4g-\gamma_a+\gamma_b+P_a\,,\quad(\mathrm{i.e.,~}g=|\Gamma_-|)\,,
\end{equation}
separates the weak and strong coupling regions. The dashed vertical
black line, specified by
\begin{equation}
  \label{eq:SunMay25231521CEST2008}
  \gamma_a=4g+\gamma_b\,,\quad(\mathrm{i.e.,~}g=|\gamma_-|)\,,
\end{equation}
corresponds to the standard criterion of SC (without incoherent
pumping).

\begin{figure}[hbpt]
  \centering
  \includegraphics[width=0.7\linewidth]{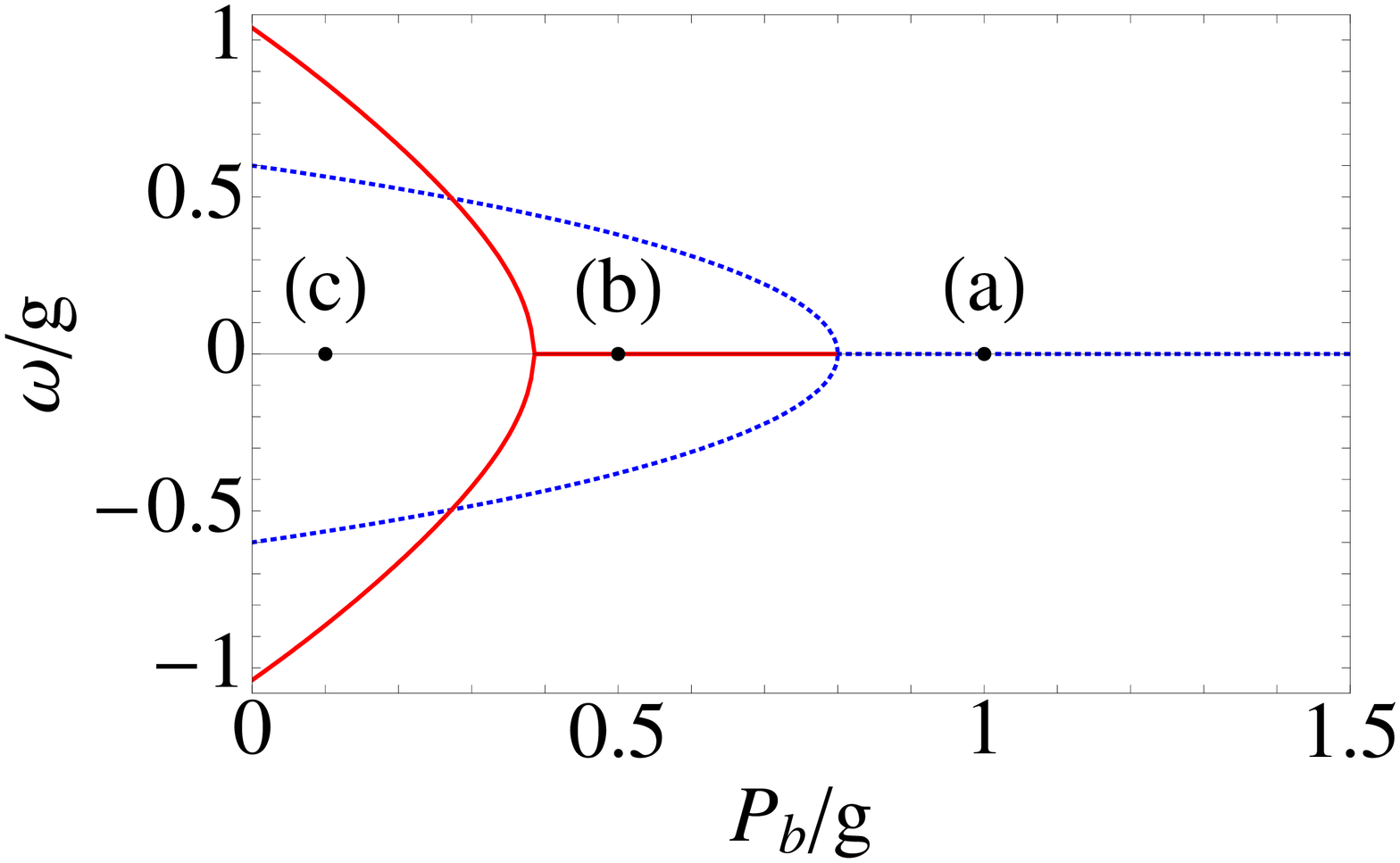}
  \caption{(Color online) Rabi slitting at resonance (dotted blue)
    $\pm R_0$, eqn~(\ref{eq:FriApr4013751UTC2008}), and the observed
    position of the peaks in the PL spectra (solid red) as a function
    of $P_b/g$.  Parameters are those of the line of points~(a), (b)
    and~(c) of Fig.~\ref{fig:FriMay16003338UTC2008}: $\gamma_a=3.8g$,
    $\gamma_b=0.1g$, $P_a=0.5g$, and the corresponding~$P_b$.}
  \label{fig:Rabi-effective}
\end{figure}

The light-blue region, labelled~1 on
Fig.~\ref{fig:FriMay16003338UTC2008}, corresponds to SC as it is
generally understood. It has a clear splitting of the lines in the
luminescence spectrum. The dark-blue region, labelled~2, corresponds
to SC, according to the requisite that~$R_0$ be real, but with a
broadening of the lines so large that in the luminescence spectrum,
eqn~(\ref{eq:spe3}), only one peak is resolved.  This region is
delimited by the brown line, which is the solution of the equation
$d^2S(\omega)/d\omega^2|_{\omega=0}=0$, i.e., with no concavity at the
origin. From this condition follows the implicit equation:
\begin{equation}
  \label{eq:MonMay19130211CEST2008}
  (3\Gamma_+-\Gamma_-)g^2+(\Gamma_--\Gamma_+)^3+g|D_0|(g^2-\Gamma_-^2+3\Gamma_+^2)=0
\end{equation}
that yields two solutions, only one of which is physical (the other is
placed on the red line~$\Gamma_+=0$, where the system diverges). Note
that this line extends into the WC region, as we shall discuss
promptly.  The distinction between line-splitting as it results from
the emergence of new dressed states in the SC, and the observation of
two peaks in the spectrum, is seen clearly on
Fig.~\ref{fig:Rabi-effective}, where the two are superimposed and seen
to differ greatly even at a qualitative level for most of the range of
parameters, coinciding only in a narrow region. The doublet as
observed in the luminescence spectrum collapse much before SC is lost.
Any estimation of system parameters, such as the coupling strength,
from a naive association of the PL spectrum with that of two-coupled
oscillators, will most likely be off by a large amount.

The last region of SC, labelled~3, is that specified
by~$4g+\gamma_b<\gamma_a<4g+\gamma_b+P_a-P_b$, i.e., that which
satisfies eqn~(\ref{eq:MonNov5033739UTC2007}) but violates
eqn~(\ref{eq:MonMay19131126CEST2008}), thereby being in SC according
to the more general definition that takes into account the effect of
the incoherent pumping, but that, according to the conventional
criterion, is in WC. For this reason, we refer to this region as of
\emph{pump-aided strong coupling}. This is a region of strong
qualitative modification of the system, that should be in WC according
to the system parameters, but that restores SC thanks to the cavity
photons forced into the system.

We now consider the other side of the blue line, that displays the
counterpart behaviours in the WC. Region~I is that of WC in its most
natural expression.  Region~II, in light, is the extension into WC of
the characteristic of featuring two maxima in the emission
spectrum. In this case, this does not correspond to a line-splitting
in the sense of SC where each peak is assigned to a renormalized
(dressed) state, but rather to a resonance of the Fano type that is
carving a hole in the single line of the weakly-coupled system. In
this region, one needs to be cautious not to read SC after the
presence of two peaks at resonance. Finally, region~III is the
counterpart of~3, in the sense that this region is in SC according to
the conventional criterion, eqn~(\ref{eq:MonMay19131126CEST2008}), for
the system parameters, when in reality the too-high electronic pumping
has bleached the strong-coupling.

\begin{figure}[hbpt]
  \centering
  \includegraphics[width=0.75\linewidth]{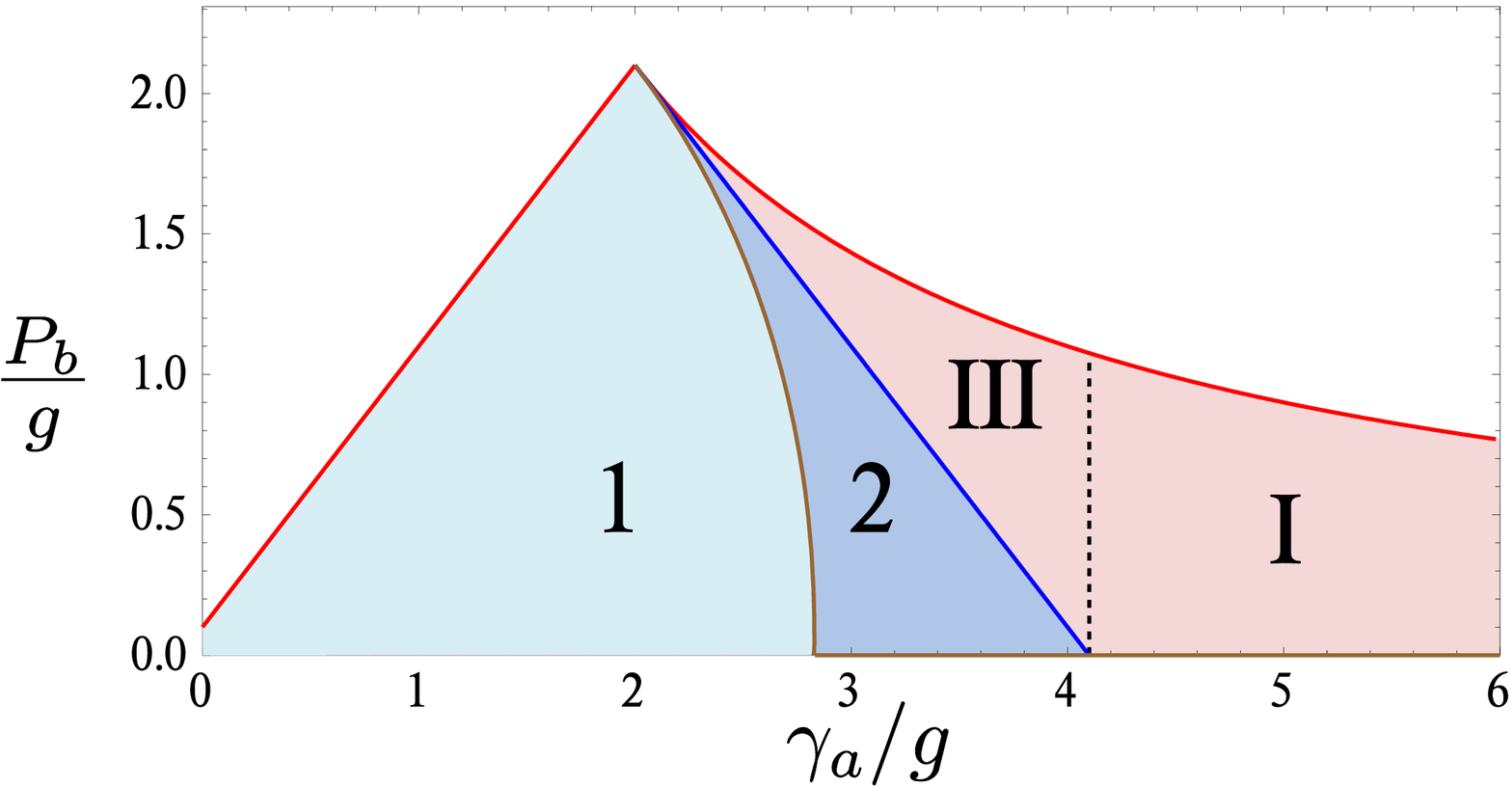}
  \caption{(Color online) The same phase space of SC/WC as in
    Fig.~\ref{fig:FriMay16003338UTC2008} as a function of $P_b/g$ and
    $\gamma_a/g$, only with $P_a/g$ now set to zero (no cavity
    pumping).  The triangle of strong coupling is displaced, and the
    regions of one peak spectra (regions 2 of SC and I of WC) are
    enlarged, following eqn~(\ref{eq:MonMay19130211CEST2008}). As a
    result regions~3 and~II have disappeared.}
  \label{fig:FriMay16004000UTC2008}
\end{figure}

In inset, we reproduce the diagram to position the five points
(a)--(e) in the various regions discussed, for which the luminescence
spectra are displayed and decomposed into their Lorentzian and
dispersive contributions, eqns~(\ref{eq:parts2}).  Case~(c), at the
lower-left angle, corresponds to SC without any pathology nor
surprise: the doublet in the luminescence spectrum---although
displaced in position as shown in Fig.~\ref{fig:Rabi-effective}---is a
faithful representation of the underlying Rabi-splitting. Increasing
pumping brings the system into region~2 where, albeit still in SC, it
does not feature a doublet anymore. The reason why, is clear on the
corresponding decomposition of the spectrum,
Fig.~\ref{fig:FriMay16003338UTC2008}(b), with a broadening of the
dressed states (in purple) too large as compared to their splitting.
Further increasing the pump brings it out of the SC region to reach
point~(a), where the two Lorentzians have collapsed. This degeneracy
of the mode emission means that the coupling only affects
perturbatively each mode. As a result, the dispersive correction has
vanished, and the spectrum now decomposes into two new Lorentzians
centred at zero, with opposite signs.

Back to point~(c), now keeping the pump constant and
increasing~$\gamma_a$, we reach point~(d). It is still in SC,
although the cavity dissipation is very large (more than four times
the coupling strength) for the small value of~$\gamma_b$ considered.
Its spectrum of emission shows, however, a clear line-splitting that
is made neatly visible thanks to the cavity (residual)
pumping~$P_a$. Increasing further the dissipation eventually brings
the system into WC, but in region~II where, again due to~$P_a\neq0$,
the spectrum remains a doublet.  In Fig.~(e), one can see, however,
that there is no Rabi splitting, and that the two peaks arise as a
result of a subtraction of the two Lorentzians centred at zero [see
the WC spectrum decomposition in
eqn~(\ref{eq:spe1weak}-\ref{eq:partWeak})]. There is probably no
need to display a spectrum from region~I, as in this case it does
not show any qualitative difference as compared to that of~(a).

The set of system parameters and the estimated effective pumping rates
allow to reconstruct Fig.~\ref{fig:FriMay16003338UTC2008} through
eqn~(\ref{eq:MonMay19125701CEST2008}--\ref{eq:MonMay19130211CEST2008}),
that contains all the physics of the system. In the following, we
shall look at variations of this representation to clarify or
illustrate those aspects that have been amply discussed before.

\begin{figure}[hbpt]
  \centering
  \includegraphics[width=0.75\linewidth]{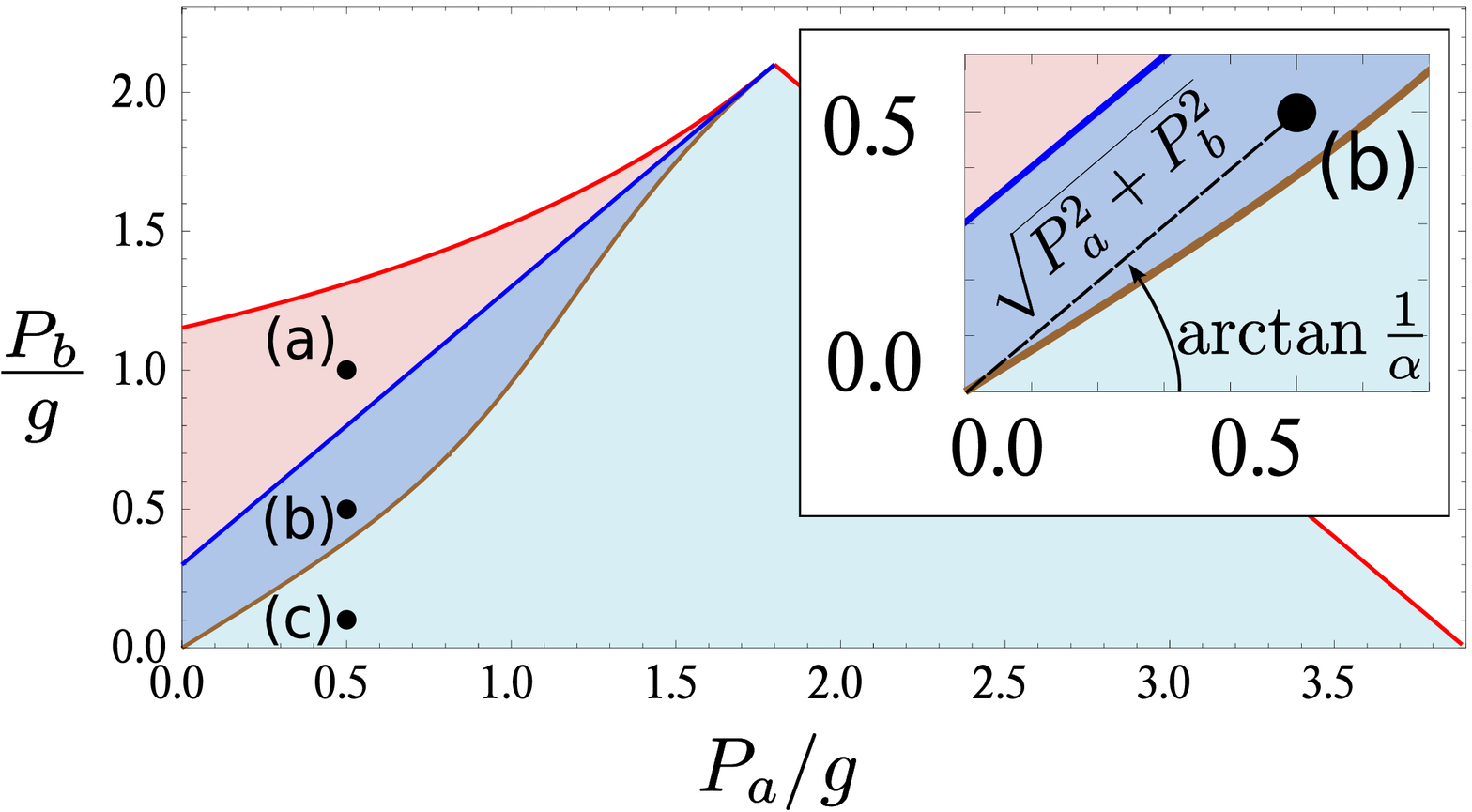}
  \caption{(Color online) Phase space of SC/WC as a function of the
    pumps $P_b/g$ and $P_a/g$ for fixed decay parameters
    $\gamma_a=3.8g$ and $\gamma_b=0.1g$. As in
    Fig.~\ref{fig:FriMay16003338UTC2008}, the red lines mark the
    physical regions and the blue one the SC (blue shades)/WC (red
    shades) transition, with the same regions~1 and~2 of SC and~III
    of WC, also with the points (a), (b) and~(c), of
    Fig.~\ref{fig:FriMay16003338UTC2008}. In inset, zoom of the
    low-pump region, showing the importance of both the angle,
    $\arctan{(1/\alpha)}$, and the magnitude, $\sqrt{P_a^2+P_b^2}$, of
    a given point.}
  \label{fig:SpeCohpump}
\end{figure}

Fig.~\ref{fig:FriMay16004000UTC2008} shows the same diagram as that of
Fig.~\ref{fig:FriMay16003338UTC2008}, only with~$P_a$ now set to zero,
i.e., corresponding to the case of a very clean sample with no
spurious QDs other than the SC coupled one, that experiences only an
electronic pumping. Observe how, as a consequence, region~3 of SC
and~II of WC have disappeared. The former was indeed the result of the
residual cavity photons helping SC. The ``pathology'' in WC of
featuring two peaks at resonance in WC has also disappeared, but most
importantly, see how region~2 has considerably increased inside the
``triangle'' of strong-coupling, meaning that the parameters required
so that the line-splitting can still be resolved in the luminescence
spectrum now put much higher demands on the quality of the structure.
This difficulty, especially in the region where~$P_b\ll g$ follows
from the ``effective quantum state in the steady state'', that we have
already discussed.  The presence of a cavity pumping, even if it is so
small that no field-intensity effects are accounted for, can favour SC
by making it visible, indeed by merely providing a photon-like
character to the quantum state. This is the manifestation in a SS of
the same influence that was observed in the SE: the luminescence
spectrum of a photon as an initial state of the coupled system, is
more visible than that of an exciton, keeping all parameters otherwise
the same (see Fig.~\ref{fig:Spe1}). Another useful picture to
highlight this last point, is that where the various regions are
plotted in terms of the pumping rates, $P_a$ and $P_b$ [see
Fig.~\ref{fig:SpeCohpump}, for the case~$\gamma_a=3.8g$ of the line
(a)--(c) in Fig.~(\ref{fig:FriMay16003338UTC2008})]. The angle of a
given point with the horizontal, linked to~$\alpha^{-1}=P_b/P_a$,
defines the exciton-like or photon-like character of the SS
established in the system, and thus determines the visibility of the
double-peak structure of SC. This is, at low pumpings, independent of
the magnitude~$\sqrt{P_a^2+P_b^2}$, as the brown line defined by
eqn~(\ref{eq:MonMay19130211CEST2008}) is approximately linear in this
region. This shows the importance of a careful determination of the
quantum state that is established in the SS by the interplay of the
pumping and decay rates, through eqn~(\ref{eq:M6}). The magnitude, on
the other hand, affects the splitting $R_0$, and the linewidths
$\Gamma_+$. In order to have a noticeable renormalization, the pumps
must be comparable to the decays.  On the one hand, the Rabi frequency
can be affected in different ways by the pumpings, depending on the
parameters.  If~$\Gamma_a=\Gamma_b$, there is, in general, no effect
of decoherence on the splitting, showing that in this case there is a
perfect symmetric coupling of the modes into the new eigenstates
(although the broadening can be large and spoil the resolution of the
Rabi splitting anyway). If they are different, for example in the
common situation that $\gamma_a-\gamma_b>P_a-P_b$, the Rabi increases
with increasing $P_a-P_b$.  On the other hand, the linewidth
$\Gamma_+=(\gamma_a+\gamma_b-P_a-P_b)/4$ presents clear bosonic
characteristics: it increases with the decays but narrows with
pumping~\cite{scully_book02a}. The intensity of the pumps also affects
the total intensity of the spectra, that is proportional to
$n^\mathrm{SS}_a$ through $\gamma_a$ and the integration time of the
apparatus. Here, however, we have focused on the normalised spectra
(i.e., the lineshape).

\section{Summary and Conclusions}
\label{sec:ThuApr3161611CEST2008}

In conclusion, we have brought under a unified formalism the
zero-dimensional light-matter interaction, both in the Weak (WC) and
Strong (SC) Coupling, for the two cases of Spontaneous Emission (SE)
of an initial state, and emission under a Steady State (SS) maintained
by an incoherent, continuous pumping. While the SE case for some
particular initial conditions (excited state of the atom) and
configurations (resonance, direct emission) has been a pillar of SC in
cQED of atoms in cavities, the extensions that we provided here to
include a continuous and incoherent pumping, are suitable to describe
the recent field of cQED in semiconductor heterostructures. Together,
they merge into an elegant and complementary theoretical edifice.

The main results of this papers are to be found in
eqns~(\ref{eq:spe1})--(\ref{eq:D}) that provide the analytical
expression for the cavity emission spectra of a system whose
specificities---such as whether it corresponds to SE or the SS
established by an incoherent continuous pumping---are provided by a
parameter~$D$, which, in the first (SE) case, is given by
eqn~(\ref{eq:Dcoh}), and in the second (SS), by eqn~(\ref{eq:M3b}).
These formulas, that allow for an arbitrary detuning between the bare
modes, reduce to more self-contained expressions at resonance, namely
eqn~(\ref{eq:speCoh1}) for SE and eqn~(\ref{eq:spe5}) for SS. The
resonance case allows an unambiguous definition of SC, depending on
whether the complex Rabi frequency,
eqn~(\ref{eq:ThuApr3185957CEST2008}), is pure imaginary (WC) or real
(SC). This corresponds, in turn, to a damping or to sustained
oscillations of the time-autocorrelation of the fields. This is
completely independent of the dynamics of the populations. SC is
characterised by the emergence of new eigenstates, with different
energies, whereas in WC, the energies remain degenerate. There is no,
however, one-to-one mapping of this splitting of the energies with the
lines observed in the luminescence spectrum. All cases can arise: one
or two peaks can be observed at resonance both in WC and SC.  In the
SC case, one peak only is observed when the energy splitting is too
small as compared to the broadening of the lines, whereas in the WC,
two peaks are seen as a result of a resonance carving a hole in a
single line, giving the illusion of a doublet (and indeed of an
anticrossing when detuning is varied). For that reason, an
understanding of the general picture is required to be able to
position a particular experiment in the space of parameters, as was
done in Fig.~\ref{fig:FriMay16003338UTC2008} and
Figs.~\ref{fig:FriMay16004000UTC2008}--\ref{fig:SpeCohpump}, rather
than to rely on a qualitative effect of anticrossing at resonance.
Figure~\ref{fig:Rabi-effective} shows how loosely related are the
observed line-splitting in the luminescence spectrum (solid red) and
the actual energy splitting of the polariton modes (dressed states, in
dotted blue). The various situations that may arise are illustrated
and discussed in Fig.~\ref{fig:FriMay16003338UTC2008}. The respective
effects of the angle and the distance to origin in the $P_a$, $P_b$
parameter-space, is shown in Fig.~\ref{fig:SpeCohpump}: the angle
accounts for the effective quantum state, that imparts on the
visibility of the splitting in the spectrum, while the magnitude
accounts for bosonic effects like line-narrowing and field-intensity
renormalization of the Rabi splitting.

This work addresses the case of bosonic excitons, that corresponds to
the case of vanishing excitations, even for genuine two-level
excitons.  As such, it also contains a lot of the physics of ground
state quantum wells excitons in planar cavities, although in this
case, another type of pumping---a so-called cross-terms pumping---is
more adequate, as particles are injected directly into the ground
state by scattering of polaritons.  On the contrary, in our present
scheme, the excitation is in terms of the bare modes, through
phonon-assisted scattering of electron-hole pairs into the QD for the
electronic pumping, $P_b$, or via Purcell-emission of weakly-coupled
spectator dots into the cavity mode for the cavity pumping, $P_a$. A
natural extension of this work is to consider fermionic excitons, that
do not admit more than one particle. In this case, the equations of
motions for the correlators are not closed, and only semi-analytical
results are available. The structure of the spectra---that in the
bosonic case decompose as a Lorentzian and a dispersive line for each
peak---becomes that of an infinite series of lines, tightly grouped
together, to give rise to multiplet structures with no more than four
peaks, albeit in a wide variety of different configurations. This
fermionic case corresponds to mapping the Jaynes-Cummings ladder to an
exact luminescence spectrum, in much the same way that we have been
doing with the Rabi doublet in this text.  This case that is of
crucial importance for the study of nonlinearity of genuine
(two-levels) QDs in semiconductor microcavities, is the subject of
part~II of this text.

\begin{acknowledgments}
  This work has been supported by the Spanish MEC under contracts QOIT
  Consolider-CSD2006-0019, MAT2005-01388 and NAN2004-09109-C04-3 and
  by CAM under contract S-0505/ESP-0200. EdV acknowledges the FPU
  scholarship (Spanish MEC).
\end{acknowledgments}

\appendix

\section{Time-resolved dynamics in the SE case}
\label{MonMay19214851CEST2008}

From eqn~(\ref{eq:M6}), the mean photon population for the SE of a
system whose initial conditions are given by
eqn~(\ref{eq:FriApr4021006UTC2008}), is:
\begin{align}
  \label{eq:ap1a}
  n_a(t)=e^{-2 \gamma_+ t}\Bigg\{&\Big[\cos{(\sqrt{2} z_1 t)}+\cosh{(\sqrt{2} z_2 t})\Big]\frac{n_a^0}{2}\\
  -&\Big[\cos{(\sqrt{2} z_1 t)}-\cosh{(\sqrt{2} z_2 t)}\Big]\frac{(\frac{\Delta^2}{4}+\gamma_-^2)n_a^0 + g^2 n_b^0+ g \Delta \Re{n_{ab}^0}-2g \gamma_- \Im{n_{ab}^0}}{2|R|^2}\nonumber\\
  +&\left[\frac{\sin{(\sqrt{2} z_1 t)}}{\sqrt{2}z_1}+\frac{\sinh{(\sqrt{2} z_2 t})}{\sqrt{2} z_2} \right]\left( g \Im{n_{ab}^0}-\gamma_- n_a^0 \right)\nonumber\\
  +&\left[\frac{\sin{(\sqrt{2}
        z_1t)}}{\sqrt{2}z_1}-\frac{\sinh{(\sqrt{2} z_2
        t})}{\sqrt{2}z_2}
  \right]\times\\&\qquad\frac{\gamma_-(\frac{\Delta^2}{4}+\gamma_-^2-g^2)n_a^0+g\Delta
    \gamma_-\Re{n_{ab}^0}+g(\frac{\Delta^2}{4}-\gamma_-^2+g^2)
    \Im{n_{ab}^0}}{|R|^2}\Bigg\}\nonumber
\end{align}

The expression for~$n_b(t)$ follows from~$a\leftrightarrow b$. The
crossed mean value that reflects the coherent coupling reads:

\begin{align}
  \label{eq:ap1b}
  &n_{ab}(t)=e^{-2 \gamma_+ t}\Bigg\{\Big[\cos{(\sqrt{2} z_1 t)}+\cosh{(\sqrt{2} z_2 t})\Big]\frac{n_{ab}^0}{2}\\
  -&\Big[\cos{(\sqrt{2} z_1 t)}-\cosh{(\sqrt{2} z_2 t)}\Big]\frac{g(\frac{\Delta}{2}+i\gamma_-) n_a^0 -g(\frac{\Delta}{2}-i\gamma_-) n_b^0-(\frac{\Delta^2}{4}+\gamma_-^2) n_{ab}^0+g^2(n_{ab}^0)^*}{2|R|^2}\nonumber\\
  +&\left[\frac{\sin{(\sqrt{2} z_1 t)}}{\sqrt{2}z_1}+\frac{\sinh{(\sqrt{2} z_2 t})}{\sqrt{2} z_2} \right]\frac{i(\Delta n_{ab}^0-g(n_a^0-n_b^0))}{2}\nonumber\\
  +&\left[\frac{\sin{(\sqrt{2} z_1
        t)}}{\sqrt{2}z_1}-\frac{\sinh{(\sqrt{2} z_2 t})}{\sqrt{2}
      z_2}
  \right]\times\\
  &\qquad\frac{g(\Delta \gamma_--i(\frac{\Delta}{2}+g^2-\gamma_-^2))
    n_a^0+g(\Delta \gamma_-+i(\frac{\Delta}{2}+g^2-\gamma_-^2))
    n_b^0+i\Delta(\frac{\Delta^2}{4}+\gamma_-^2+g^2)
    n_{ab}^0}{2|R|^2}\Bigg\}\nonumber
\end{align}
where we have defined the complex parameters:
\begin{align}
  \label{eq:ap2}
  z_{1,2}=&\sqrt{|R|^2\pm (g^2-\gamma_-^2+\frac{\Delta^2}{4})}\,.
\end{align}

In case of SC, at resonance, ($g>|\gamma_-|$), both $z_{1,2}$ are
real, due to the fact that $|R|^2>|\gamma_-\Delta|$ for whatever the
detuning, and they can be written in a more transparent way:
\begin{align}
  \label{eq:ap3}
  z_{1,2}=|R|\sqrt{1\pm\sqrt{1-\left(\frac{\gamma_-\Delta}{|R|^2}\right)^2}}\,.
\end{align}
The mean value $n_a(t)$ is plotted in Fig.~\ref{fig:Spe1}(b) and
\ref{fig:Spe12}(c) for SC and WC, respectively, for the parameters in
the caption.

It might be of interest to note that
eqns~(\ref{eq:ap1a}--\ref{eq:ap1b}), but also eqn~(\ref{eq:M5b}) in
the SE case, and therefore all the results that follow from them, are
reproduced by introducing decay as an imaginary part to the energies
in the Heisenberg picture, i.e., upgrading $\omega_{a,b}$
to~$\omega_{a,b}-i\gamma_{a,b}/2$ and solving directly in a full
Hamiltonian picture the operator equations of motion:~$\partial_t
c(t)=i[H,c(t)]$ with~$c=a,b$. This simple expedient resurges plainly
for instance in eqn~(\ref{eq:ThuApr3185957CEST2008}), that is an exact
and important result derived with the full dissipative formalism, but
which follows directly from eqn~(\ref{eq:ThuJun5115514BST2008}) with
complex energies. This recourse to complex energies is however not
valid in the Schrödinger picture and/or with the pumping terms.

\section{Derivation of the luminescence spectrum in the SS from the
  general expression}
\label{MonMay26103617CEST2008}

Expression~(\ref{eq:ThuApr3171338CEST2008}) is general, but its final
form in term of a time-integrated Fourier transforms of~$G^{(1)}$,
eqn~(\ref{eq:ThuApr3172523CEST2008}), is not rigorous for the SS case
since both the numerator and the denominator are infinite
quantities. Their ratio, however, produces a finite quantity, which
bears all the characteristic of the optical emission spectrum in a~SS,
and---may be more convincingly---recovers the well-established and
rigorously derived Wiener-Khintchine formula,
eqn~(\ref{eq:MonMay19214206CEST2008}). Note that this formula is,
strictly speaking, an arcane mathematical result in the theory of
stochastic processes. There, $S(\omega)$ is a mesure of the strength
of the fluctuations of the Fourier component at
frequency~$\omega$~\cite{mandel95a}. It has no strict connection with
a physical signal, as both infinite negative and positive times are
required for its demonstration, which violates causality among other
things. For a rigorous and extended discussion of a physical optical
spectrum, see the excellent discussion by Eberly and
W\'odkiezicw~\cite{eberly77a}. As the step from the general
eqn~(\ref{eq:ThuApr3171338CEST2008}) to the SS,
eqn~(\ref{eq:MonMay19214206CEST2008}), can be made straightforwardly,
we provide it below: it merely consists in cancelling the diverging
quantities by obtaining the final result as a limit of the
time-integrated spectrum. We believe that the starting point being so
clear mathematically, and this presentation so satisfying in its
simultaneous coverage of both the SE and SS cases, it is worth
considering for a pedagogical purpose.

The population and the autocorrelator can be decomposed as a transient
(TR) and steady-state (SS) values:
\begin{subequations}
  \label{eq:S4}
  \begin{align}
    \langle\ud{a}a\rangle(t)&=\langle\ud{a}a\rangle^\mathrm{TR}(t)+\lim_{t\rightarrow\infty}\langle\ud{a}a\rangle(t)\,,\\
    G^{(1)}(t,\tau)&=\langle\ud{a}(t)a(t+\tau)\rangle^\mathrm{TR}+\lim_{t\rightarrow\infty}\langle\ud{a}(t)a(t+\tau)\rangle\,,
  \end{align}
\end{subequations}
where~$\lim_{t\rightarrow\infty}\langle\ud{a}a\rangle(t)=n_a^\mathrm{SS}$,
cf.~eqn~(\ref{eq:MonMay19124027CEST2008}). We rewrite
eqn~(\ref{eq:ThuApr3172523CEST2008}) as the time integration of the
Fourier transform until time~$T$, that is left to increase without
bounds:
\begin{equation}
  \label{eq:TueMay20214306UTC2008}
  S(\omega)=\lim_{T\rightarrow\infty}\frac{1}{\pi\int_{0}^T\langle\ud{a}a\rangle(t)dt}\Re\int_{t=0}^T\int_{\tau=0}^\infty G^{(1)}(t,\tau)e^{i\omega\tau}\,d\tau dt\,.
\end{equation}
Substituting eqns~(\ref{eq:S4}) in this expression, we can keep track
of the terms that cancel (one can check, from the explicit results of
the text, the convergence of the quantities
$N\equiv\int_0^\infty\langle\ud{a}a\rangle^\mathrm{TR}(t)dt$
and~$M_\tau=\int_0^\infty\langle\ud{a}(t)a(t+\tau)\rangle^\mathrm{TR}e^{i\omega\tau}\,dt$,
for all~$\tau>0$):
\begin{equation}
  \label{eq:S5}
  S^\mathrm{SS}(\omega)=\frac{1}{\pi}\lim_{T\rightarrow\infty}\frac{1}{N+Tn_a^\mathrm{SS}}
  \Re\int_{\tau=0}^{\infty}\bigg[M_\tau
  +T\lim_{t\rightarrow\infty}\langle\ud{a}(t)a(t+\tau)\rangle\bigg]e^{i\omega\tau}d\tau\,.
\end{equation}
Since the norm of the Fourier transform of~$M_\tau$ is also bounded
(that, again, can be checked from the explicit result), the limit
in~$T$ yields:
\begin{equation}
  \label{eq:MonMay26103804CEST2008}
  S^\mathrm{SS}(\omega)=\frac{1}{\pi}\frac{1}{n_a^\mathrm{SS}}\lim_{t\rightarrow\infty}\Re\int_{0}^{\infty}\langle\ud{a}(t)a(t+\tau)\rangle e^{i\omega\tau}d\tau\,,
\end{equation}
which is the formula of the text,
eqn~(\ref{eq:MonMay19214206CEST2008}).

\bibliography{Sci,books,psip}

\end{document}